\documentclass[]{aiaa-tc}

\usepackage{amsmath}
\usepackage{amssymb}
\usepackage{bm}
\usepackage{mathrsfs}
\usepackage{verbatim}
\usepackage{subfigure}
\usepackage{placeins}
\usepackage{overpic}
\usepackage{multicol}
\usepackage{color}
\usepackage[table]{xcolor}

\usepackage[pdftex]{hyperref}
\hypersetup{
pdfauthor   = {},
pdftitle    = {},
pdfsubject  = {},
pdfkeywords = {},
pdfcreator  = {PDFLaTeX with hyperref package},
pdfproducer = {pdflatex},
colorlinks  = true,
linkcolor   = blue,
anchorcolor = red,
citecolor   = blue,
filecolor   = red,
urlcolor    = red
}

\title{Computational Performance of a LES Solver for Supersonic Jet Flow Applications}

\author{
Carlos Junqueira-Junior\thanks{Postdoctoral Research Fellow, 
Aerodynamics Division, Departamento de Ci\^{e}ncia e Tecnologia 
Aeroespacial, DCTA/IAE/ALA;
E-mail: junior.hmg@gmail.com.} \hspace{0.2 cm} and \hspace{0.2 cm}
Jo\~{a}o Luiz F. Azevedo\thanks{Senior Research Engineer, Aerodynamics Division,
Departamento de Ci\^{e}ncia e Tecnologia Aeroespacial, DCTA/IAE/ALA;
E-mail: joaoluiz.azevedo@gmail.com. AIAA Fellow.}\\
{\normalsize\itshape Instituto de Aeron\'{a}utica e Espa\c{c}o, 12228-904
S\~{a}o Jos\'{e} dos Campos, SP, Brazil}\\
\and
Sami Yamouni\thanks{Postdoctoral Reasearch Fellow,
Graduate Program on Computer Sciences and Electrical Engineering,
Departamento de Ci\^{e}ncia e Tecnologia Aeroespacial, DCTA/ITA;
E-mail: sami.yamouni@gmail.com.}\\
{\normalsize\itshape Instituto Tecnol\'{o}gico de Aeron\'{a}utica, 12228-900
S\~{a}o Jos\'{e} dos Campos, SP, Brazil}\\
\and
William R. Wolf
\thanks{Assistant Professor, Faculty of Mechanical Engineering;
E-mail: wolf@fem.unicamp.br. AIAA Member}\\
{\normalsize\itshape Universidade Estadual de Campinas, 13083-970
Campinas, SP, Brazil}\\
}

 \AIAApapernumber{2015-NUMBER}
 \AIAAconference{22nd AIAA Computational Fluid Dynamic Conference, June, 2015, Dallas, Texas}
 \AIAAcopyright{\AIAAcopyrightD{2015}}

\begin{document}

\maketitle


\begin{abstract}

\section*{Abstract}

An in-house large eddy simulation tool is developed in order to 
reproduce high fidelity results of compressible jet flows. 
The large eddy simulation formulation is written using the finite 
difference approach, with an explicit time integration and using a 
second order spatial discretization. The energy equation is carefully 
discretized in order to model the energy equation of the filtered 
Navier-Stokes formulation. Such nu\-me\-ri\-cal studies are very 
expensive and demand high performance computing. Message passage 
interface protocols are implemented into the code in order to perform 
parallel computations. The present work addresses the computational 
performance of the solver running on up to 400 processors in parallel. 
Different mesh configurations, whose size varies from approximately 
5.9 million points to approximately 1.0 billion points, are evaluate 
in the current paper. Speedup and efficiency curves are evaluated 
in order to assess the strong scalability of the solver.

\end{abstract}




\section{Introduction}

Solid structure of different parts of launch vehicles and 
experimental apparatus on board can be da\-ma\-ged during the 
take off and also during the transonic flight of such vehicles
due to vibrational acoustic stress resulted from pressure 
fluctuations. Such fluctuations are originated from the 
complex interaction between the high-temperature/high-velocity 
exhaustion gases from the rocket engines. The acoustic design 
constraints of launch vehicles have encouraged the studies of 
aeroacoustic fields around compressible jet flows for aerospace 
applications. Instituto de Aeronautica e Espa\c{c}o (IAE) has 
been using large eddy simulations (LES) \cite{Junior16,jr16-aiaa}
coupled with the Ffowcs Williams and Hawkings approach 
\cite{Wolf2012} in order to study the aeroacoustic of supersonic 
jet flow configurations. The LES studies are very expensive in the 
computational context and strongly demand parallel computing. The 
present work addresses the computational performance of the solver 
using up to 400 processors in parallel. The speedup and computational 
efficiency of the solver are measured using different mesh and 
partition configurations and different number of computational cores.

JAZzY \cite{Junior16} is the LES solver which is used in the present 
work. It is an in-house computational tool developed regarding the 
study of unsteady turbulent supersonic jet flow configurations. 
The formulation is written using the finite difference approach. 
Inviscid numerical fluxes are calculated using a second order 
accurate centered scheme with the explicit addition of artificial 
dissipation. A five steps second order accurate Runge-Kutta is 
the chosen time marching method. A formulation based on the System 
I set of equations \cite{Vreman1995} is used here in order to 
model the filtered terms of the energy equation. 
Numerical simulation of perfectly expanded jets are performed and 
compared with numerical \cite{Mendez10} and experimental 
\cite{bridges2008turbulence} data.

The code is written using the FORTRAN 90 standards. It uses the {\it
HDF5} \cite{folk99,folk11} and the {\it CGNS} \cite{Poirier98,
Poirier00,legensky02} libraries for the I/O operations. The mesh
is partitioned into the axial and azimuthal directions. Two 
layer ghost points are created at the surroundings of the local domain
in order to carry neighbor partitions data. The data exchange between 
partitions is performed through message passing interface (MPI) 
protocols \cite{Dongarra95}.

Simulations of a perfectly expanded jet are performed using different 
mesh partitioning and different number of processors in order to evaluate 
the computational performance of the code. In the present nine meshes are
are studied running on up to 400 processors in parallel. The size of the mesh 
starts with 5.8 million points and scales to 1.0 billion points. The speedup 
and computational efficiency curves are presented and compared in order to 
study the strong scalability of the code.


  \section{Large Eddy Simulation Filtering}

The large eddy simulation is based on the principle 
of scale separation, which is addressed as a 
filtering procedure in a mathematical formalism. A 
modified version of the the System I filtering 
approach \cite{Vreman1995} is used in present work
which is given by
\begin{equation}
\begin{array}{c}
\displaystyle \frac{\partial \overline{\rho} }{\partial t} + \frac{\partial}{\partial x_{j}} 
\left( \overline{\rho} \widetilde{  u_{j} } \right) = 0 \, \mbox{,}\\
\displaystyle \frac{\partial}{\partial t} \left( \overline{ \rho } \widetilde{ u_{i} } \right) 
+ \frac{\partial}{\partial x_{j}} 
\left( \overline{ \rho } \widetilde{ u_{i} } \widetilde{ u_{j} } \right)
+ \frac{\partial \overline{p}}{\partial x_{i}} 
- \frac{\partial {\tau}_{ij}}{\partial x_{j}}  
+ \frac{1}{3} \frac{\partial}{\partial x_{j}}\left( {\delta}_{ij} \sigma_{ii}\right)
= 0 \, \mbox{,} \\ 
\displaystyle \frac{\partial \overline{e}}{\partial t} 
+ \frac{\partial}{\partial x_{j}} 
\left[ \left( \overline{e} + \overline{p} \right)\widetilde{u_{j}} \right]
- \frac{\partial}{\partial x_{j}}\left({\tau}_{ij} \widetilde{u_{i}} \right)
+ \frac{1}{3} \frac{\partial}{\partial x_{j}}
\left[ \left( \delta_{ij}{\sigma}_{ii} \right) \widetilde{u_{i}} \right]
+ \frac{\partial {q}_{j}}{\partial x_{j}} = 0 \, \mbox{,}
\end{array}
\label{eq:modified_system_I}
\end{equation}
in which $t$ and $x_{i}$ are independent variables 
representing time and spatial coordinates of a 
Cartesian coordinate system $\textbf{x}$, respectively. 
The components of the velocity vector $\textbf{u}$ are 
written as $u_{i}$, and $i=1,2,3$. Density, pressure and 
total energy per mass unit are denoted by $\rho$, $p$ and 
$e$, respectively. The $\left( \overline{\cdot} \right)$ and
$\left( \tilde{\cdot} \right)$ operators are used in order 
to represent filtered and Favre averaged properties, 
respectively. The System I formulation neglects the double 
correlation term and the total energy per mass unit is written 
as 
\begin{equation}
	\overline{e} = \frac{\overline{p}}{\gamma - 1} 
	+ \frac{1}{2} \rho \widetilde{u}_{i} \widetilde{u}_{i} \, \mbox{.} 
\end{equation}
The heat flux, $q_{j}$, is given by
\begin{equation}
	{q}_{j} = \left(\kappa+{\kappa}_{sgs}\right) 
	\frac{\partial \widetilde{T}}{\partial x_{j}} 
	\, \mbox{.}
	\label{eq:q_mod}
\end{equation}
where $T$ is the static temperature and $\kappa$ is the thermal 
conductivity, which can by expressed by
\begin{equation}
\kappa = \frac{\mu C_{p}}{Pr} \, \mbox{,}
\end{equation}
The thermal conductivity is a function of the specific heat at 
constant pressure, $Cp$, of the Prandtl number, $Pr$, which is 
equal to $0.72$ for air, and of the dynamic viscosity, $\mu$.
The SGS thermal conductivity, $\kappa_{sgs}$, is written as
\begin{equation}
	\kappa_{sgs} = \frac{\mu_{sgs} C_{p}}{ {Pr}_{sgs} } 
	\, \mbox{,}
	\label{eq:kappa_sgs}
\end{equation}
where ${Pr}_{sgs}$ is the SGS Prandtl number, which is
equal to $0.9$ for static SGS models and $\mu_{sgs}$
is the eddy viscosity which is calculated by the SGS
closure. The dynamic viscosity, $\mu$, can be calculated 
using the Sutherland law,
\begin{eqnarray}
	\mu \left( \widetilde{T} \right) = \mu_{\infty} 
	\left( \frac{\widetilde{T}}{\widetilde{T}_{\infty}}
	\right)^{\frac{3}{2}} 
	\frac{\widetilde{T}_{0}+S_{1}}{\widetilde{T}+S_{1}} &
\mbox{with} \: S_{1} = 110.4K \, \mbox{ .}
\label{eq:sutherland}
\end{eqnarray}
Density, static pressure and static temperature are correlated 
by the equation of state given by
\begin{equation}
	\overline{p} = \rho R \widetilde{T} \, \mbox{,}
\end{equation}
where $R$ is the gas constant, written as
\begin{equation}
R = C_{p} - C_{v} \, \mbox{,}
\end{equation}
and $C_{v}$ is the specific heat at constant volume.
The shear-stress tensor, $\tau_{ij}$, is written 
according to the Stokes hypothesis and includes
the eddy viscosity, $\mu_{sgs}$,
\begin{equation}
	{\tau}_{ij} = 2 \left(\mu+{\mu}_{sgs}\right) 
	\left( \tilde{S}_{ij} - \frac{1}{3} \delta_{ij} \tilde{S}_{kk} \right) \,
	\label{eq:tau_mod}
\end{equation}
in which $\tilde{S}_{ij}$, components of rate-of-strain tensor, are 
given by
\begin{equation}
	\tilde{S}_{ij} = \frac{1}{2} 
	\left( \frac{\partial \tilde{u}_{i}}{\partial x_{j}} 
	+ \frac{\partial \tilde{u}_{j}}{\partial x_{i}} 
\right) \, \mbox{.}
\end{equation}
The SGS stress tensor components are written using the eddy 
viscosity \cite{Sagaut05},
\begin{equation}
    \sigma_{ij} = - 2 \mu_{sgs} \left( \tilde{S}_{ij} 
	- \frac{1}{3} \tilde{S}_{kk} \right)
    + \frac{1}{3} \delta_{ij} \sigma_{kk}
    \, \mbox{.}
    \label{eq:sgs_visc}
\end{equation}
The eddy viscosity, $\mu_{sgs}$, and the components of the 
isotropic part of the SGS stress tensor, $\sigma_{kk}$, are
modeled by the SGS closure.


\section{Subgrid Scale Modeling}

The theoretical formulation of subgrid scales closures included in the present work
is discussed in the present section. The closure models presented here are founded 
on the homogeneous turbulence theory, which is usually developed in the spectral 
space as an attempt to quantify the interaction between the different scales of 
turbulence.

\subsection{Smagorinky Model}

The Smagorinsky model \cite{Smagorinsky63} is one of the simplest algebric models 
for the deviatory part of the SGS tensor used in large-eddy simulations. The 
isotropic part of the SGS tensor is neglected for Smagorinsky model in the
current work. This SGS closure is a classical model based the large scales 
properties and is written as
\begin{equation}
\mu_{sgs} = \left( \rho C_{s} \Delta \right)^{2} | \widetilde{S} | \, \mbox{,}
\end{equation}
where
\begin{equation}
| \tilde{S} | = \left( 2 \tilde{S}_{ij} \tilde{S}_{ij} \right)^{\frac{1}{2}} \, \mbox{,}
\end{equation}
$\Delta$ is the filter size and $C_{s}$ is the Smagorinsky constant. 
Several attempts can be found in the literature regarding the 
evaluation of the Smagorinsky constant. The value of this constant 
is adjusted to improve the results of different flow 
configurations. In pratical terms, the Smagorinsky subgrid 
model has a flow dependency of the constant which takes value 
ranging from 0.1 to 0.2 depending on the flow. The suggestion 
of Lilly \cite{Lilly67}, $C_{s}=0.148$, is used in the current 
work.


This model is generally over-dissipative in regions of large mean strain. 
This is particularly true in the transitional region between laminar and 
turbulent flows. Moreover, the limiting behavior near the wall is not
correct, and the model predictions correlate poorly with the exact subgrid 
scale tensor \cite{Garnier09}. However, it is a very simple model and, 
with the use of damping function and good calibration, can be successfully 
applied on large-eddy simulations.

\subsection{Vreman Model}

Vreman \cite{vreman2004} proposed a turbulence model that can correctly predict
inhomogeneous turbulent flows. For such flows, the eddy viscosity should become 
small in laminar and transitional regions. This requirement is unfortunately not 
satisfied by existing simple eddy-viscosity closures such as the classic 
Smagorinsky model \cite{Smagorinsky63,Lilly65,Deardorff70}. The Vreman SGS model
is very simple and is given by
\begin{equation}
	\mu_{sgs} = \rho \, \bm{c} \, 
	\sqrt{\frac{B_{\beta}}{\alpha_{ij} \alpha_{ij}}} 
	\,\mbox{,}
\end{equation}
with
\begin{equation}
	\alpha_{ij} = \frac{\partial \tilde{u}_{j}}{\partial x_{i}} 
	\, \mbox{,}
\end{equation}
\begin{equation}
	\beta_{ij} = \Delta^{2}_{m}\alpha_{mi}\alpha_{mj}
\end{equation}
and 
\begin{equation}
	B_{\beta} = \beta_{11}\beta_{22} - \beta_{12}^{2} 
	          + \beta_{11}\beta_{33} - \beta_{13}^{2}
			  + \beta_{22}\beta_{33} - \beta_{23}^{2}
			  \, \mbox{.}
\end{equation}
The constant $\bm{c}$ is related to the Smagorinsky constant, $C_{s}$, 
and it is given by
\begin{equation}
	\bm{c} = 2.5 \, C_{s}^{2} 
	\, \mbox{,}
\end{equation}
and $\Delta_{m}$ is the filter width in each direction. In the present work,
the isotropic part of the SGS tensor is neglected for the Vreman model.
The $\alpha$ symbol represents the matrix of first order derivatives of the 
filtered components of velocity, $\tilde{u}_{i}$. The SGS eddy-viscosity is 
defined as zero when $\alpha_{ij}\alpha_{ij}$ equals zero. Vreman
\cite{vreman2004} affirms that the tensor $\beta$ is proportional to the 
gradient model \cite{Leonard74,Clark79} in its general anisotropic form
\cite{vreman1996}.

The Vreman model can be classified as very simple model because it is expressed 
in first-order derivatives and it dos not involves explicit filtering, averaging, 
clipping procedures and is rotationally invariant for isotropic filter widths. The 
model is originally created for incompressible flows and it has presented good 
results for two incompressible flows configurations: the transitional and turbulent 
mixing layer at high Reynolds number and the turbulent channel flow \cite{vreman1996}. 
In both cases, the Vreman model is found to be more accurate than the classical 
Smagorinsky model and as good as the dynamic Smagorinsky model.

\subsection{Dynamic Smagorinsky Model}

Germano {\it et al.} \cite{germano90} developed a dynamic SGS model in order to 
overcome the issues of the classical Smagorinsky closure. The model uses the strain 
rate fields at two different scales and thus extracts spectral information in 
the large-scale field to extrapolate the small stresses \cite{moin91}. 
The coefficients of the model are computed instantaneously in the dynamic model. 
They are function of the positioning in space and time rather than being specified 
a priori. Moin {\it et al.} \cite{moin91} extended the work of Germano for compressible 
flows. The dynamic Smagorinsky model for compressible flow configurations is detailed in 
the present section.

The Dynamic model introduces the test filter, $\widehat{\left( \cdot \right)}$, which
has a larger filter width, $\widehat{\Delta}$, than the one of the resolved grid filter, 
$\overline{\left( \cdot \right)}$. The use of test filters generates a second field 
with larger scales than the resolved field. The Yoshizawa model \cite{Yoshizawa86} is 
used for the isotropic portion of the SGS tensor and it is written as
\begin{equation}
	\sigma_{ll} = 2 C_{I} \overline{\rho} {\Delta}^{2}|\tilde{S}|^{2}
	\, \mbox{,}
	\label{eq:yoshizawa}
\end{equation}
where $C_{I}$ is defined by
\begin{equation}
	C_{I} = \frac{\biggl\langle \widehat{\overline{\rho} \tilde{u}_{l} \tilde{u}_{l}} -
	        \left( \widehat{\overline{\rho}\tilde{u}_{l}}
	               \widehat{\overline{\rho}\tilde{u}_{l}}/
				   \widehat{\overline{\rho}} \right)\biggr\rangle }
		   {\biggl\langle 2 \widehat{\Delta}^{2} \widehat{\overline{\rho}}
		   |\widehat{\overline{S}}|^{2} - 
	        2 {\Delta}^{2} \widehat{ \overline{\rho}
		   |\overline{S}|^{2}}\biggr\rangle}
		   \, \mbox{.}
		   \label{eq:av_ci}
\end{equation}
A volume averaging, here indicated by $\langle \,\, \rangle$, is suggest by 
Moin {\it et al} \cite{moin91} and by Garnier {\it et al} in order to avoid 
numerical issues. The eddy viscosity, $\mu_{sgs}$, is calculated using the 
same approach used by static Smagorinsky model,
\begin{equation}
\mu_{sgs} = \left( \rho C_{ds} \Delta \right)^{2} | \tilde{S} | \, \mbox{,}
\end{equation}
where
\begin{equation}
| \tilde{S} | = \left( 2 \tilde{S}_{ij} \tilde{S}_{ij} \right)^{\frac{1}{2}} 
\, \mbox{,}
\end{equation}
and $C_{ds}$ is the dynamic constant of the model, which is given by
\begin{equation}
	C_{ds} = \frac{\biggl\langle
	        \left[ \widehat{\overline{\rho} \tilde{u}_{i} \tilde{u}_{j}} -
	        \left( \widehat{\overline{\rho}\tilde{u}_{i}}
	        \widehat{\overline{\rho}\tilde{u}_{j}}/
			\widehat{\overline{\rho}} \right) \right]\tilde{S}_{ij} - 
            \frac{1}{3}\tilde{S}_{mm}
	        \left(\mathscr{T}_{ll} - \widehat{\sigma}_{ll}\right)
			\biggr\rangle}{\biggl\langle
			2 {\Delta}^{2}\left[
			\widehat{\overline{\rho}|\tilde{S}|\tilde{S}_{ij}}\tilde{S}_{ij}
			- \frac{1}{3}\left(\overline{\rho}|\tilde{S}|\tilde{S}_{mm}\right)^{\widehat{ }}
			\tilde{S}_{ll}\right] -
            2 \widehat{\Delta}^{2}\left(
			\widehat{\overline{\rho}}|\widehat{\tilde{S}}|
			\widehat{\tilde{S}}_{ij}\tilde{S}_{ij} -
            \frac{1}{3}\widehat{\overline{\rho}}|\widehat{\tilde{S}}|
            \widehat{\tilde{S}}_{mm} \tilde{S}_{ll}\right)
			\biggr\rangle}
	\, \mbox{.}
    \label{eq:av_cd}
\end{equation}
The SGS Prandtl number is computed using the dynamic constant, $C_{ds}$, and 
written as
\begin{equation}
	{Pr}_{sgs} = C_{ds} \frac{\biggl\langle \Delta^{2} 
	\biggl(\overline{\rho}|\tilde{S}|\frac{\partial \overset{\sim}{T}}{\partial x_{j}}
	\biggr)^{\widehat{ }} \,
	\frac{\partial\overset{\sim}{T}}{\partial x_{j}} -
	\widehat{\Delta}^{2}\widehat{\overline{\rho}}|\widehat{\tilde{S}}|
	\frac{\partial \overset{\sim}{T}}{\partial x_{j}}
	\frac{\partial \overset{\sim}{T}}{\partial x_{j}} \biggr\rangle}
	{\biggl\langle\left[ \widehat{\overline{\rho} \tilde{u}_{j} \overset{\sim}{T}} - 
	\left( \widehat{\overline{\rho} \tilde{u_{j}}} 
	\widehat{\overline{\rho} \overset{\sim}{T}} \right)/
		   \widehat{\overline{\rho}}\right] 
		   \frac{\partial \overset{\sim}{T}}{\partial x_{j}}\biggr\rangle}
	\, \mbox{.}
	\label{eq:av_Pr_sgs}
\end{equation}
%



  \section{Transformation of Coordinates}

The formulation is written in the a general curvilinear coordinate 
system in order to facilitate the implementation and add more 
generality for the CFD tool. Hence, the filtered Navier-Stokes 
equations can be written in strong conservation form for a 
3-D general curvilinear coordinate system as
\begin{equation}
	\frac{\partial \hat{Q}}{\partial t} 
	+ \frac{\partial }{\partial \xi}\left(\hat{\mathbf{E}_{e}}-\hat{\mathbf{E}_{v}}\right) 
	+ \frac{\partial}{\partial \eta}\left(\hat{\mathbf{F}_{e}}-\hat{\mathbf{F}_{v}}\right)
	+ \frac{\partial}{\partial \zeta}\left(\hat{\mathbf{G}_{e}}-\hat{\mathbf{G}_{v}}\right) 
	= 0 \, \mbox{.}
	\label{eq:vec-LES}
\end{equation}
In the present work, the chosen general coordinate transformation is given by
\begin{eqnarray}
	\xi & = & \xi \left(x,y,z,t \right)  \, \mbox{,} \nonumber\\
	\eta & = & \eta \left(x,y,z,t \right)  \, \mbox{,} \\
	\zeta & = & \zeta \left(x,y,z,t \right)\ \, \mbox{.} \nonumber
\end{eqnarray}
In the jet flow configuration, $\xi$ is the axial jet flow direction, $\eta$ is 
the radial direction and $\zeta$ is the azimuthal direction. The vector of
conserved properties is written as
\begin{equation}
	\hat{Q} = J^{-1} \left[ \overline{\rho} \quad \overline{\rho}\tilde{u} \quad 
	\overline{\rho}\tilde{v} \quad \overline{\rho}\tilde{w} \quad \overline{e} \right]^{T} 
	\quad \mbox{,}
	\label{eq:hat_Q_vec}
\end{equation}
where the Jacobian of the transformation, $J$, is given by
\begin{equation}
	J = \left( x_{\xi} y_{\eta} z_{\zeta} + x_{\eta}y_{\zeta}z_{\xi} +
	           x_{\zeta} y_{\xi} z_{\eta} - x_{\xi}y_{\zeta}z_{\eta} -
			   x_{\eta} y_{\xi} z_{\zeta} - x_{\zeta}y_{\eta}z_{\xi} 
	    \right)^{-1} \, \mbox{,}
\end{equation}
and
\begin{eqnarray}
	\displaystyle x_{\xi}   = \frac{\partial x}{\partial \xi}  \, \mbox{,} & 
	\displaystyle x_{\eta}  = \frac{\partial x}{\partial \eta} \, \mbox{,} & 
	\displaystyle x_{\zeta} = \frac{\partial x}{\partial \zeta}\, \mbox{,} \nonumber \\
	\displaystyle y_{\xi}   = \frac{\partial y}{\partial \xi}  \, \mbox{,} & 
	\displaystyle y_{\eta}  = \frac{\partial y}{\partial \eta} \, \mbox{,} & 
	\displaystyle y_{\zeta} = \frac{\partial y}{\partial \zeta}\, \mbox{,} \\
	\displaystyle z_{\xi}   = \frac{\partial z}{\partial \xi}  \, \mbox{,} & 
	\displaystyle z_{\eta}  = \frac{\partial z}{\partial \eta} \, \mbox{,} & 
	\displaystyle z_{\zeta} = \frac{\partial z}{\partial \zeta}\, \mbox{.} \nonumber
\end{eqnarray}

The inviscid flux vectors, $\hat{\mathbf{E}}_{e}$, $\hat{\mathbf{F}}_{e}$ and 
$\hat{\mathbf{G}}_{e}$, are given by
{\small
\begin{eqnarray}
	\hat{\mathbf{E}}_{e} = J^{-1} \left\{\begin{array}{c}
		\overline{\rho} U \\
		\overline{\rho}\tilde{u} U + \overline{p} \xi_{x} \\
		\overline{\rho}\tilde{v} U + \overline{p} \xi_{y} \\
		\overline{\rho}\tilde{w} U + \overline{p} \xi_{z} \\
		\left( \overline{e} + \overline{p} \right) U - \overline{p} \xi_{t}
\end{array}\right\} \, \mbox{,} &
%
	\hat{\mathbf{F}}_{e} = J^{-1} \left\{\begin{array}{c}
		\overline{\rho} V \\
		\overline{\rho}\tilde{u} V + \overline{p} \eta_{x} \\
		\overline{\rho}\tilde{v} V + \overline{p} \eta_{y} \\
		\overline{\rho}\tilde{w} V + \overline{p} \eta_{z} \\
		\left( \overline{e} + \overline{p} \right) V - \overline{p} \eta_{t}
\end{array}\right\} \, \mbox{,} &
%
	\hat{\mathbf{G}}_{e} = J^{-1} \left\{\begin{array}{c}
		\overline{\rho} W \\
		\overline{\rho}\tilde{u} W + \overline{p} \zeta_{x} \\
		\overline{\rho}\tilde{v} W + \overline{p} \zeta_{y} \\
		\overline{\rho}\tilde{w} W + \overline{p} \zeta_{z} \\
		\left( \overline{e} + \overline{p} \right) W - \overline{p} \zeta_{t}
	\end{array}\right\} \, \mbox{.}
	\label{eq:hat-flux-G}
\end{eqnarray}
}
The contravariant velocity components, $U$, $V$ and $W$, are calculated as
%
%
\begin{eqnarray}
  U = \xi_{x}\overline{u} + \xi_{y}\overline{v} + \xi_{z}\overline{w} 
  \, \mbox{,} \nonumber \\
  V = \eta_{x}\overline{u} + \eta_{y}\overline{v} + \eta_{z}\overline{w} 
  \, \mbox{,} \\
  W = \zeta_{x}\overline{u} + \zeta_{y}\overline{v} + \zeta_{z}\overline{w} 
  \, \mbox{.} \nonumber
  \label{eq:vel_contrv}
\end{eqnarray}
The metric terms are given by
%
%
\begin{eqnarray}
	\xi_{x} = J \left( y_{\eta}z_{\zeta} - y_{\zeta}z_{\eta} \right) \, \mbox{,} & 
	\xi_{y} = J \left( z_{\eta}x_{\zeta} - z_{\zeta}x_{\eta} \right) \, \mbox{,} & 
	\xi_{z} = J \left( x_{\eta}y_{\zeta} - x_{\zeta}y_{\eta} \right) \, \mbox{,} \nonumber \\
	\eta_{x} = J \left( y_{\eta}z_{\xi} - y_{\xi}z_{\eta} \right) \, \mbox{,} & 
	\eta_{y} = J \left( z_{\eta}x_{\xi} - z_{\xi}x_{\eta} \right) \, \mbox{,} & 
	\eta_{z} = J \left( x_{\eta}y_{\xi} - x_{\xi}y_{\eta} \right) \, \mbox{,} \\
	\zeta_{x} = J \left( y_{\xi}z_{\eta} - y_{\eta}z_{\xi} \right) \, \mbox{,} & 
	\zeta_{y} = J \left( z_{\xi}x_{\eta} - z_{\eta}x_{\xi} \right) \, \mbox{,} & 
	\zeta_{z} = J \left( x_{\xi}y_{\eta} - x_{\eta}y_{\xi} \right) \, \mbox{.} \nonumber \\
\end{eqnarray}

The viscous flux vectors, $\hat{\mathbf{E}}_{v}$, $\hat{\mathbf{F}}_{v}$ and 
$\hat{\mathbf{G}}_{v}$, are written as
\begin{equation}
	\hat{\mathbf{E}}_{v} = J^{-1} \left\{\begin{array}{c}
		0 \\
		\xi_{x}{\tau}_{xx} +  \xi_{y}{\tau}_{xy} + \xi_{z}{\tau}_{xz} \\
		\xi_{x}{\tau}_{xy} +  \xi_{y}{\tau}_{yy} + \xi_{z}{\tau}_{yz} \\
		\xi_{x}{\tau}_{xz} +  \xi_{y}{\tau}_{yz} + \xi_{z}{\tau}_{zz} \\
		\xi_{x}{\beta}_{x} +  \xi_{y}{\beta}_{y} + \xi_{z}{\beta}_{z} 
	\end{array}\right\} \, \mbox{,}
	\label{eq:hat-flux-Ev}
\end{equation}
\begin{equation}
	\hat{\mathbf{F}}_{v} = J^{-1} \left\{\begin{array}{c}
		0 \\
		\eta_{x}{\tau}_{xx} +  \eta_{y}{\tau}_{xy} + \eta_{z}{\tau}_{xz} \\
		\eta_{x}{\tau}_{xy} +  \eta_{y}{\tau}_{yy} + \eta_{z}{\tau}_{yz} \\
		\eta_{x}{\tau}_{xz} +  \eta_{y}{\tau}_{yz} + \eta_{z}{\tau}_{zz} \\
		\eta_{x}{\beta}_{x} +  \eta_{y}{\beta}_{y} + \eta_{z}{\beta}_{z} 
	\end{array}\right\} \, \mbox{,}
	\label{eq:hat-flux-Fv}
\end{equation}
\begin{equation}
	\hat{\mathbf{G}}_{v} = J^{-1} \left\{\begin{array}{c}
		0 \\
		\zeta_{x}{\tau}_{xx} +  \zeta_{y}{\tau}_{xy} + \zeta_{z}{\tau}_{xz} \\
		\zeta_{x}{\tau}_{xy} +  \zeta_{y}{\tau}_{yy} + \zeta_{z}{\tau}_{yz} \\
		\zeta_{x}{\tau}_{xz} +  \zeta_{y}{\tau}_{yz} + \zeta_{z}{\tau}_{zz} \\
		\zeta_{x}{\beta}_{x} +  \zeta_{y}{\beta}_{y} + \zeta_{z}{\beta}_{z} 
	\end{array}\right\} \, \mbox{,}
	\label{eq:hat-flux-Gv}
\end{equation}
where $\beta_{x}$, $\beta_{y}$ and $\beta_{z}$ are defined as
\begin{eqnarray}
	\beta_{x} = {\tau}_{xx}\tilde{u} + {\tau}_{xy}\tilde{v} +
	{\tau}_{xz}\tilde{w} - \overline{q}_{x} \, \mbox{,} \nonumber \\
	\beta_{y} = {\tau}_{xy}\tilde{u} + {\tau}_{yy}\tilde{v} +
	{\tau}_{yz}\tilde{w} - \overline{q}_{y} \, \mbox{,} \\
	\beta_{z} = {\tau}_{xz}\tilde{u} + {\tau}_{yz}\tilde{v} +
	{\tau}_{zz}\tilde{w} - \overline{q}_{z} \mbox{.} \nonumber
\end{eqnarray}
%

\section{Dimensionless Formulation}

A convenient nondimensionalization is necessary in to order to achieve a consistent 
implementation of the governing equations of motion. Dimensionless formulation 
yields to a more general numerical tool. There is no need to change the formulation 
for each configuration intended to be simulated. Moreover, dimensionless formulation 
scales all the necessary properties to the same order of magnitude which is a 
computational advantage \cite{BIGA02}. Dimensionless variables are presented in the 
present section in order perform the nondimensionalization of Eq.\ 
\eqref{eq:vec-LES}

The dimensionless time, $\underline{t}$, is written as function of the 
speed of sound of the jet at the inlet, $a_{j}$, and of a reference lenght, $l$,
\begin{equation}
	\underline{t} = t \frac{a_{j}}{l} \, \mbox{.}
	\label{eq:non-dim-time}
\end{equation}
%
%
The dimensionless velocity components are obtained using the speed of sound of the 
jet at the inlet,
\begin{equation}
	\underline{\textbf{u}} = \frac{\textbf{u}}{a_{j}} \, \mbox{.}
	\label{eq:non-dim-vel}
\end{equation}
Dimensionless pressure and energy are calculated using density and speed of the sound
of the jet at the inlet as
\begin{equation}
	\underline{p} = \frac{p}{\rho_{j}a_{j}^{2}} \, \mbox{,}
	\label{eq:non-dim-press}
\end{equation}
\begin{equation}
	\underline{E} = \frac{E}{\rho_{j}a_{j}^{2}} \, \mbox{.}
	\label{eq:non-dim-energy}
\end{equation}
Dimensionless density, $\underline{\rho}$, temperature, $\underline{T}$ and 
viscosity, $\underline{\mu}$, are calculated using freestream properties
\begin{equation}
	\underline{\rho} = \frac{\rho}{\rho_{j}} \, \mbox{.}
	\label{eq:non-dim-rho}
\end{equation}

One can use the dimensionless properties described above in order to write the 
dimensionless form of the RANS equations as
\begin{equation}
	\frac{\partial \underline{Q}}{\partial t} + 
	\frac{\partial \underline{\mathbf{E}}_{e}}{\partial \xi} +
	\frac{\partial \underline{\mathbf{F}}_{e}}{\partial \eta} + 
	\frac{\partial \underline{\mathbf{G}}_{e}}{\partial \zeta} =
	\frac{M_{j}}{Re} \left( \frac{\partial \underline{\mathbf{E}}_{v}}{\partial \xi} 
	+ \frac{\partial \underline{\mathbf{F}}_{v}}{\partial \eta} 
	+ \frac{\partial \underline{\mathbf{G}}_{v}}{\partial \zeta} \right)	\, \mbox{,}
	\label{eq:vec-underline-split-RANS}
\end{equation}
where the underlined terms are calculated using dimensionless properties.
The Mach number of the jet, $M_{j}$, and the Reynolds number are based on 
the mean inlet velocity of the jet, $U_{j}$, diamenter of the inlet, $D$,
and freestream properties such as speed of sound, $a_{\infty}$, density, 
$\rho_{\infty}$ and viscosity, $\mu_{\infty}$,
\begin{eqnarray}
	M_{j}=\frac{U_j}{a_{\infty}} & \mbox{and} & 
	Re = \frac{\rho_{j}U_{j}D}{\mu_{j}} \, \mbox{.}
\end{eqnarray}

\section{Numerical Formulation}

The governing equations previously described are discretized in a 
structured finite difference context for general curvilinear 
coordinate system \cite{BIGA02}. The numerical flux is calculated 
through a central difference scheme with the explicit addition 
of the anisotropic scalar artificial dissipation of Turkel and Vatsa
\cite{Turkel_Vatsa_1994}. The time integration is performed by an 
explicit, 2nd-order, 5-stage Runge-Kutta scheme 
\cite{jameson_mavriplis_86, Jameson81}.  Conserved properties
and artificial dissipation terms are properly treated near boundaries in order
to assure the physical correctness of the numerical formulation. 
 
\subsection{Spatial Discretization}

For the sake of simplicity the formulation discussed in the present section
is no longer written using bars. However, the reader should notice that the 
equations are dimensionless and filtered. The Navier-Stokes equations, 
presented in Eq.\ \eqref{eq:vec-underline-split-RANS}, are discretized in 
space in a finite difference fashion and, then, rewritten as
\begin{equation}
	\left(\frac{\partial Q}{\partial t}\right)_{i,j,k} \  
	= \  - RHS_{i,j,k} \, \mbox{,}	
	\label{eq:spatial_discret}
\end{equation}
where $RHS$ is the right hand side of the equation and it is written as function of 
the numerical flux vectors at the interfaces between grid points,
\begin{eqnarray}
	{RHS}_{i,j,k} & = & 
	\frac{1}{\Delta \xi} \left( 
	{\mathbf{E}_{e}}_{(i+\frac{1}{2},j,k)} - {\mathbf{E}_{e}}_{(i-\frac{1}{2},j,k)} - 
	{\mathbf{E}_{v}}_{(i+\frac{1}{2},j,k)} + {\mathbf{E}_{v}}_{(i-\frac{1}{2},j,k)} 
	\right) \nonumber \\
	& & \frac{1}{\Delta \eta} \left( 
	{\mathbf{F}_{e}}_{(i,j+\frac{1}{2},k)} - {\mathbf{F}_{e}}_{(i,j-\frac{1}{2},k)} - 
	{\mathbf{F}_{v}}_{(i,j+\frac{1}{2},k)} + {\mathbf{F}_{v}}_{(i,j-\frac{1}{2},k)} 
	\right) \\
	& & \frac{1}{\Delta \zeta} \left( 
	{\mathbf{G}_{e}}_{(i,j,k+\frac{1}{2})} - {\mathbf{G}_{e}}_{(i,j,k-\frac{1}{2})} - 
	{\mathbf{G}_{v}}_{(i,j,k+\frac{1}{2})} + {\mathbf{G}_{v}}_{(i,j,k-\frac{1}{2})} 
	\right) \, \mbox{.} \nonumber
\end{eqnarray}
For the general curvilinear coordinate case 
$\Delta \xi = \Delta \eta = \Delta \zeta = 1$. The anisotropic scalar 
artificial dissipation method of Turkel and Vatsa \cite{Turkel_Vatsa_1994}
is implemented through the modification of the inviscid flux vectors, 
$\mathbf{E}_{e}$, $\mathbf{F}_{e}$ and $\mathbf{G}_{e}$. The numerical scheme 
is nonlinear and allows the selection between artificial dissipation terms of 
second and fourth differences, which is very important for capturing discontinuities 
in the flow. The numerical fluxes are calculated at interfaces in order to reduce 
the size of the calculation cell and, therefore, facilitate the implementation 
of second derivatives since the the concept of numerical fluxes vectors is 
used for flux differencing. Only internal interfaces receive the corresponding 
artificial dissipation terms, and differences of the viscous flux vectors 
use two neighboring points of the interface. 

The inviscid flux vectors, with the addition of the artificial dissipation
contribution, can be written as
\begin{eqnarray}
	{\mathbf{E}_{e}}_{(i \pm \frac{1}{2},j,k)} 
	= \frac{1}{2} \left( {\mathbf{E}_{e}}_{(i,j,k)} + {\mathbf{E}_{e}}_{(i \pm 1,j,k)} \right)
	- J^{-1} \mathbf{d}_{(i \pm \frac{1}{2},j,k)} \, \mbox{,} \nonumber \\
	{\mathbf{F}_{e}}_{(i,j\pm \frac{1}{2},k)} 
	= \frac{1}{2} \left( {\mathbf{F}_{e}}_{(i,j,k)} + {\mathbf{F}_{e}}_{(i,j \pm 1,k)} \right)
	- J^{-1} \mathbf{d}_{(i,j \pm \frac{1}{2},k)} \, \mbox{,} \label{eq:inv_flux_vec}\\
	{\mathbf{G}_{e}}_{(i,j,k\pm \frac{1}{2})} 
	= \frac{1}{2} \left( {\mathbf{G}_{e}}_{(i,j,k)} + {\mathbf{G}_{e}}_{(i,j,k \pm 1)} \right)
	- J^{-1} \mathbf{d}_{(i,j,k \pm \frac{1}{2})} \, \mbox{,} \nonumber
\end{eqnarray}
in which the $\mathbf{d}_{(i\pm 1,j,k)}$,$\mathbf{d}_{(i,j\pm 1,k)}$ and $\mathbf{d}_{(i,j,k\pm 1)}$ terms
are the Turkel and Vatsa \cite{Turkel_Vatsa_1994} artificial dissipation terms
in the $i$, $j$, and $k$ directions respectively. The scaling of the artificial
dissipation operator in each coordinate direction is weighted by its own spectral 
radius of the corresponding flux Jacobian matrix, which gives the non-isotropic 
characteristics of the method \cite{BIGA02}. The artificial dissipation contribution
in the $\xi$ direction is given by
\begin{eqnarray}
	\mathbf{d}_{(i + \frac{1}{2},j,k)} & = & 
	\lambda_{(i + \frac{1}{2},j,k)} \left[ \epsilon_{(i + \frac{1}{2},j,k)}^{(2)}
	\left( \mathcal{W}_{(i+1,j,k)} - \mathcal{W}_{(i,j,k)} \right) \right. \label{eq:dissip_term}\\
	& & \epsilon_{(i + \frac{1}{2},j,k)}^{(4)} \left( \mathcal{W}_{(i+2,j,k)} 
	- 3 \mathcal{W}_{(i+1,j,k)} + 3 \mathcal{W}_{(i,j,k)} 
	- \mathcal{W}_{(i-1,j,k)} \right) \left. \right] \, \mbox{,} \nonumber
\end{eqnarray}
in which
\begin{eqnarray}
	\epsilon_{(i + \frac{1}{2},j,k)}^{(2)} & = &
	k^{(2)} \mbox{max} \left( \nu_{(i+1,j,k)}^{d}, 
	\nu_{(i,j,k)}^{d} \right) \, \mbox{,} \label{eq:eps_2_dissip} \\
	\epsilon_{(i + \frac{1}{2},j,k)}^{(4)} & = &
	\mbox{max} \left[ 0, k^{(4)} - \epsilon_{(i + \frac{1}{2},j,k)}^{(2)} \right] 
	\, \mbox{.} \label{eq:eps_4_dissip}
\end{eqnarray}
The original article \cite{Turkel_Vatsa_1994} recomends using $k^{(2)}=0.25$ and 
$k^{(4)}=0.016$ for the dissipation artificial constants. The pressure 
gradient sensor, $\nu_{(i,j,k)}^{d}$, for the $\xi$ direction is written as
\begin{equation}
	\nu_{(i,j,k)}^{d} = \frac{|p_{(i+1,j,k)} - 2 p_{(i,j,k)} + p_{(i-1,j,k)}|}
	                          {p_{(i+1,j,k)} - 2 p_{(i,j,k)} + p_{(i-1,j,k)}} 
	\, \mbox{.}
\label{eq:p_grad_sensor}
\end{equation}
The $\mathcal{W}$ vector from Eq.\ \eqref{eq:dissip_term} is calculated as a function of the
conserved variable vector, $\hat{Q}$, written in Eq.\ \eqref{eq:hat_Q_vec}.
The formulation intends to keep the total enthalpy constant in the final converged 
solution, which is the correct result for the Navier-Stokes equations with 
$Re \rightarrow \infty$. This approach is also valid for the viscous formulation 
because the dissipation terms are added to the inviscid flux terms, in which they 
are really necessary to avoid nonlinear instabilities of the numerical formulation. 
The $\mathcal{W}$ vector is given by
\begin{equation}
	\mathcal{W} = \hat{Q} + \left[0 \,\, 0 \,\, 0 \,\, 0 \,\, p \right]^{T} \, \mbox{.}
	\label{eq:W_dissip}
\end{equation}
The spectral radius-based scaling factor, $\lambda$, for the $i-\mbox{th}$ 
direction is written
\begin{equation}
	\lambda_{(i+\frac{1}{2},j,k)} = \frac{1}{2} \left[ 
	\left( \overline{\lambda_{\xi}}\right)_{(i,j,k)} + 
	\left( \overline{\lambda_{\xi}}\right)_{(i+1,j,k)}
	\right] \, \mbox{,} 
\end{equation}
where
\begin{equation}
    \overline{\lambda_{\xi}}_{(i,j,k)} = \lambda_{\xi} \left[ 1 + 
	\left(\frac{\lambda_{\eta}}{\lambda_{\xi}} \right)^{0.5} + 
	\left(\frac{\lambda_{\zeta}}{\lambda_{\xi}} \right)^{0.5} \right] 
	\, \mbox{.}
\end{equation}
The spectral radii, $\lambda_{\xi}$, $\lambda_{\eta}$ and $\lambda_{\zeta}$ are given
by
\begin{eqnarray}
	\lambda_{\xi} & = & 
	|U| + a \sqrt{\xi_{x}^{2} + \eta_{y}^{2} + \zeta_{z}^{2}} 
	\, \mbox{,} \nonumber \\
	\lambda_{\xi} & = & 
	|V| + a \sqrt{\xi_{x}^{2} + \eta_{y}^{2} + \zeta_{z}^{2}} 
	\, \mbox{,} \\
	\lambda_{\xi} & = & 
	|W| + a \sqrt{\xi_{x}^{2} + \eta_{y}^{2} + \zeta_{z}^{2}} 
	\, \mbox{,} \nonumber
\end{eqnarray}
in which, $U$, $V$ and $W$ are the contravariants velocities in the $\xi$, $\eta$
and $\zeta$, previously written in Eq.\ \eqref{eq:vel_contrv}, and $a$ is the local 
speed of sound, which can be written as
\begin{equation}
	a = \sqrt{\frac{\gamma p}{\rho}} \, \mbox{.}
\end{equation}
The calculation of artificial dissipation terms for the other coordinate directions
are completely similar and, therefore, they are not discussed in the present work.

\subsection{Time Marching Method}

The time marching method used in the present work is a 2nd-order, 5-step Runge-Kutta
scheme based on the work of Jameson \cite{Jameson81, jameson_mavriplis_86}. 
The time integration can be written as
\begin{equation}
	\begin{array}{ccccc}
	Q_{(i,jk,)}^{(0)} & = & Q_{(i,jk,)}^{(n)} \, \mbox{,} & & \\
	Q_{(i,jk,)}^{(l)} & = & Q_{(i,jk,)}^{(0)} -  
	& \alpha_{l} {\Delta t}_{(i,j,k)} {RHS}_{(i,j,k)}^{(l-1)} \, & 
	\,\,\,\, l = 1,2 \cdots 5, \\
	Q_{(i,jk,)}^{(n+1)} & = & Q_{(i,jk,)}^{(5)} \, \mbox{,} & &
	\end{array}
	\label{eq:localdt}
\end{equation}
in which $\Delta t$ is the time step and $n$ and $n+1$ indicate the property
values at the current and at the next time step, respectively. The literature
\cite{Jameson81, jameson_mavriplis_86} recommends 
\begin{equation}
	\begin{array}{ccccc}
		\alpha_{1} = \frac{1}{4} \,\mbox{,} & \alpha_{2} = \frac{1}{6} \,\mbox{,} &
		\alpha_{3} = \frac{3}{8} \,\mbox{,} & \alpha_{4} = \frac{1}{2} \,\mbox{,} & 
		\alpha_{5} = 1 \,\mbox{,} 
	\end{array}
\end{equation}
in order to improve the numerical stability of the time integration. The present
scheme is theoretically stable for $CFL \leq 2\sqrt{2}$, under a linear analysis
\cite{BIGA02}.

\section{Boundary Conditions} \label{sec:BC}

\begin{figure}[ht]
       \begin{center}
		   \subfigure[Lateral view of boundary conditions.]{
           \includegraphics[width=0.475\textwidth]
		   {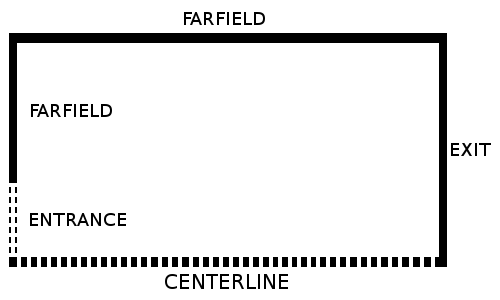} 
		   \label{fig:bc-1}
		   }
		   \subfigure[Frontal view of boundary conditions.]{
           \includegraphics[width=0.475\textwidth]
		   {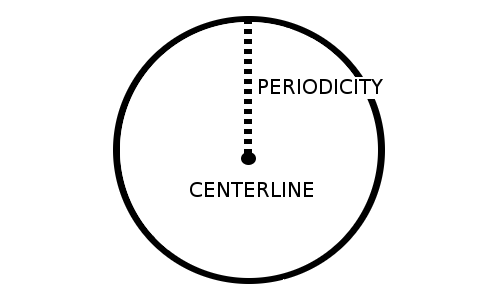} 
		   \label{fig:bc-2}
		   }
		   \caption{Lateral and frontal views of the computational domain 
		   indicating boundary conditions.}
		   \label{fig:bc}
	   \end{center}
\end{figure}

The geometry used in the present work presents a
cylindrical shape which is gererated by the rotation of 
a 2-D plan around a centerline. Figure \ref{fig:bc} 
presents a lateral view and a frontal view of the 
computational domain used in the present work and 
the positioning of the entrance, exit, centerline, 
far field and periodic boundary conditions. A discussion
on all boundary conditions is performed in the following 
subsections.

\subsection{Far Field Boundary}

Riemann invariants \cite{Long91} are used to implement far field boundary conditions.
They are derived from the characteristic relations for the Euler equations.
At the interface of the outer boundary, the following expressions apply
\begin{eqnarray}
	\mathbf{R}^{-} = {\mathbf{R}}_{\infty}^{-} & = & q_{n_\infty}-\frac{2}{\gamma-1}a_\infty\, \mbox{,}  \\
	\mathbf{R}^{+} = {\mathbf{R}}_{e}^{+} & = & q_{n_e}-\frac{2}{\gamma-1}a_e \, \mbox{,}
	\label{eq:R-farfield}
\end{eqnarray}
where $\infty$ and $e$ indexes stand for the property in the freestream and in the 
internal region, respectively. $q_n$ is the velocity component normal to the outer surface,
defined as
\begin{equation}
	q_n={\bf u} \cdot \vec{n} \, \mbox{,}
	\label{eq:qn-farfield}
\end{equation}
and $\vec{n}$ is the unit outward normal vector 
\begin{equation}
	\vec{n}=\frac{1}{\sqrt{\eta_{x}^2+\eta_{y}^2+\eta_{z}^2}}
	[\eta_x \ \eta_y \ \eta_z ]^T \, \mbox{.}
	\label{eq:norm-vec}
\end{equation}
Equation \eqref{eq:qn-farfield} assumes that the $\eta$ direction is pointing from the jet to the 
external boundary. Solving for $q_n$ and $a$, one can obtain
\begin{eqnarray}
	q_{n f} = \frac{\mathbf{R}^+ + \mathbf{R}^-}{2} \, \mbox{,} & \ & 
	a_f = \frac{\gamma-1}{4}(\mathbf{R}^+ - \mathbf{R}^-) \, \mbox{.}
	\label{eq: qn2-farfield}
\end{eqnarray}
The index $f$ is linked to the property at the boundary surface and will be used to update 
the solution at this boundary. For a subsonic exit boundary, $0<q_{n_e}/a_e<1$, the 
velocity components are derived from internal properties as
 \begin{eqnarray}
	 u_f&=&u_e+(q_{n f}-q_{n_e})\eta_x \, \mbox{,} \nonumber \\ 
	 v_f&=&v_e+(q_{n f}-q_{n_e})\eta_y \, \mbox{,} \\ 
	 w_f&=&w_e+(q_{n f}-q_{n_e})\eta_z \, \mbox{.} \nonumber
	 \label{eq:vel-farfield}
 \end{eqnarray}
Density and pressure properties are obtained by extrapolating the entropy from 
the adjacent grid node,
\begin{eqnarray}
	\rho_f = 
	\left(\frac{\rho_{e}^{\gamma}a_{f}^2}{\gamma p_e} \right)^{\frac{1}{\gamma-1}}
	\, \mbox{,} & \ &
	p_{f} = \frac{\rho_{f} a_{f}^2}{\gamma} \, \mbox{.} \nonumber
	 \label{eq:rhop-farfield}
\end{eqnarray}
For a subsonic entrance, $-1<q_{n_e}/a_e<0$, properties are obtained similarly 
from the freestream variables as
\begin{eqnarray}
	u_f&=&u_\infty+(q_{n f}-q_{n_\infty})\eta_x \, \mbox{,} \nonumber \\
	v_f&=&v_\infty+(q_{n f}-q_{n_\infty})\eta_y \, \mbox{,} \\
	w_f&=&w_\infty+(q_{n f}-q_{n_\infty})\eta_z \, \mbox{,} \nonumber
	\label{eq:vel2-farfield}
\end{eqnarray}
\begin{equation}
	\rho_f = 
	\left(\frac{\rho_{\infty}^{\gamma}a_{f}^2}{\gamma p_\infty} \right)^{\frac{1}{\gamma-1}}
	\, \mbox{.}
	\label{eq:rhop2-farfield}
\end{equation}
For a supersonic exit boundary, $q_{n_e}/a_e>1$, the properties are extrapolated 
from the interior of the domain as
\begin{eqnarray}
	\rho_f&=&\rho_e \, \mbox{,} \nonumber\\
	u_f&=&u_e \, \mbox{,} \nonumber\\
	v_f&=&v_e \, \mbox{,} \\
	w_f&=&w_e \, \mbox{,} \nonumber\\
	e_f&=&e_e \, \mbox{,} \nonumber   
	\label{eq:supso-farfield}
\end{eqnarray}
and for a supersonic entrance, $q_{n_e}/a_e<-1$, the properties are extrapolated 
from the freestream variables as
\begin{eqnarray}
	\rho_f&=&\rho_\infty \, \mbox{,}  \nonumber\\
	u_f&=&u_\infty \, \mbox{,}  \nonumber\\
	v_f&=&v_\infty \, \mbox{,} \\
	w_f&=&w_\infty \, \mbox{,} \nonumber\\
	e_f&=&e_\infty \, \mbox{.} \nonumber
	\label{eq:supso2-farfield}
\end{eqnarray}

\subsection{Entrance Boundary}

For a jet-like configuration, the entrance boundary is divided in two areas: the
jet and the area above it. The jet entrance boundary condition is implemented through 
the use of the 1-D characteristic relations for the 3-D Euler equations for a flat
velocity profile. The set of properties then determined is computed from within and 
from outside the computational domain. For the subsonic entrance, the $v$ and $w$ components
of the velocity are extrapolated by a zero-order extrapolation from inside the 
computational domain and the angle of flow entrance is assumed fixed. The rest of the properties 
are obtained as a function of the jet Mach number, which is a known variable. 
\begin{eqnarray}
	\left( u \right)_{1,j,k} & = & u_{j} \, \mbox{,} \nonumber \\
	\left( v \right)_{1,j,k} & = & \left( v \right)_{2,j,k} \,\mbox{,} \\
	\left( w \right)_{1,j,k} & = & \left( w \right)_{2,j,k} \, \mbox{.} \nonumber
	\label{eq:vel-entry}
\end{eqnarray}
The dimensionless total temperature and total pressure are defined with the isentropic relations:
\begin{eqnarray}
	T_t = 1+\frac{1}{2}(\gamma-1)M_{\infty}^{2} \, & \mbox{and} & 
	P_t = \frac{1}{\gamma}(T_t)^{\frac{\gamma}{\gamma-1}} \, \mbox{.}
	\label{eq:Tot-entry}
\end{eqnarray}
The dimensionless static temperature and pressure are deduced from Eq.\ \eqref{eq:Tot-entry},
resulting in
\begin{eqnarray}
	\left( T \right)_{1,j,k}=\frac{T_t}{1+\frac{1}{2}(\gamma-1)(u^2+v^2+w^2)_{1,j,k}} \, 
	& \mbox{and} & 
	\left( p \right)_{1,j,k}=\frac{1}{\gamma}(T)_{1,j,k}^{\frac{\gamma}{\gamma-1}} \, \mbox{.}
	\label{eq:Stat-entry}
\end{eqnarray}
For the supersonic case, all conserved variables receive jet property values.

The far field boundary conditions are implemented outside of the jet area in order to correctly
propagate information comming from the inner domain of the flow to the outter region of 
the simulation. However, in the present case, $\xi$, instead of $\eta$, as presented in 
the previous subsection, is the normal direction used to define the Riemann invariants.

\subsection{Exit Boundary Condition}

At the exit plane, the same reasoning of the jet entrance boundary is applied. This time, 
for a subsonic exit, the pressure is obtained from the outside and all other variables are 
extrapolated from the interior of the computational domain by a zero-order extrapolation. The 
conserved variables are obtained as
\begin{eqnarray}
	(\rho)_{I_{MAX},j,k} &=& \frac{(p)_{I_{MAX},j,k}}{(\gamma-1)(e)_{I_{MAX}-1,j,k}} \mbox{,} \\
	(\vec{u})_{I_{MAX},j,k} &=& (\vec{u})_{I_{MAX}-1,j,k}\mbox{,} \\
	(e_i)_{I_{MAX},j,k} &=& 
	(\rho)_{I_{MAX},j,k}\left[ (e)_{I_{MAX}-1,j,k}+
	\frac{1}{2}(\vec{u})_{I_{MAX},j,k}\cdot(\vec{u})_{I_{MAX},j,k} \right] \, \mbox{,}
	\label{eq:exit}
\end{eqnarray}
in which $I_{MAX}$ stands for the last point of the mesh in the axial direction. For 
the supersonic exit, all properties are extrapolated from the interior domain.

\subsection{Centerline Boundary Condition}

The centerline boundary is a singularity of the coordinate transformation, and, hence, 
an adequate treatment of this boundary must be provided. The conserved properties 
are extrapolated from the ajacent longitudinal plane and are averaged in the azimuthal 
direction in order to define the updated properties at the centerline of the jet.

\begin{figure}[ht]
       \begin{center}
       {\includegraphics[width=0.5\textwidth]{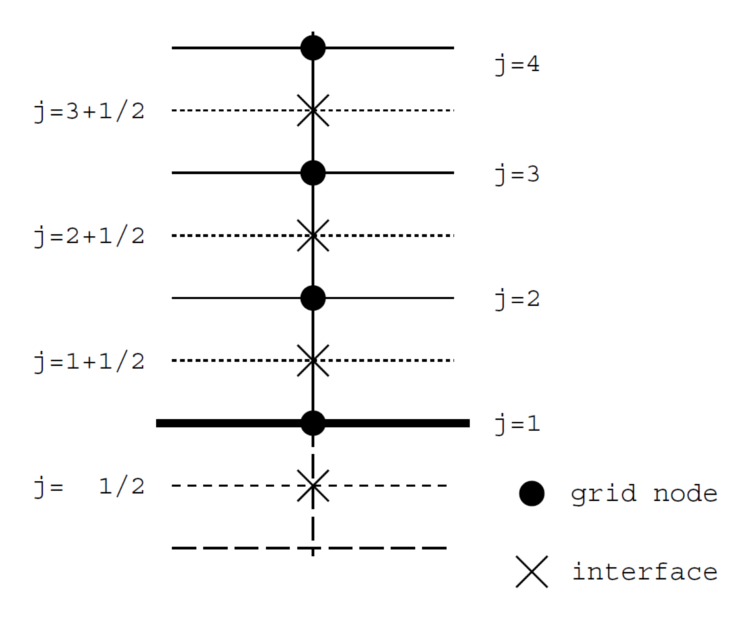}}\\
	   \caption{Boundary points dissipation \cite{BIGA02}.}\label{fig:centerline}
       \end{center}
\end{figure}

The fourth-difference terms of the artificial dissipation scheme, used in the present 
work, are carefully treated in order to avoid the five-point difference stencils at 
the centerline singularity. 
If one considers the flux balance at one grid point near the centerline boundary in 
a certain coordinate direction, let $w_{j}$ denote a component of the $\mathcal{W}$ 
vector from Eq.\ \eqref{eq:W_dissip} and $\mathbf{d}_{j}$ denote the corresponding artificial
dissipation term at the mesh point $j$. In the present example, 
$\left(\Delta w\right)_{j+\frac{1}{2}}$ stands for the difference between the solution
at the interface for the points $j+1$ and $j$. The fouth-difference of the dissipative
fluxes from Eq.\ \eqref{eq:dissip_term} can be written as
\begin{equation}
	\mathbf{d}_{j+\frac{1}{2}} = \left( \Delta w \right)_{j+\frac{3}{2}} 
	- 2 \left( \Delta w \right)_{j+\frac{1}{2}}
	+ \left( \Delta w \right)_{j-\frac{1}{2}} \, \mbox{.}
\end{equation}
Considering the centerline and the point $j=1$, as presented in Fig.\ 
\ref{fig:centerline}, the calculation of $\mathbf{d}_{1+\frac{1}{2}}$ demands the 
$\left( \Delta w \right)_{\frac{1}{2}}$ term, which is unknown since it is outside the
computation domain. In the present work a extrapolation is performed and given by
\begin{equation}
	\left( \Delta w \right)_{\frac{1}{2}} =
	- \left( \Delta w \right)_{1+\frac{1}{2}} \, \mbox{.}
\end{equation}
This extrapolation modifies the calculation of $\mathbf{d}_{1+\frac{1}{2}}$ that can be written as
\begin{equation}
	\mathbf{d}_{j+\frac{1}{2}} = \left( \Delta w \right)_{j+\frac{3}{2}} 
	- 3 \left( \Delta w \right)_{j+\frac{1}{2}} \, \mbox{.}
\end{equation}
The approach is plausible since the centerline region is smooth and does not have high
gradient of properties.

\subsection{Periodic Boundary Condition}

A periodic condition is implemented between the first ($K=1$) and the last point in the 
azimutal direction ($K=K_{MAX}$) in order to close the 3-D computational domain. There 
are no boundaries in this direction, since all the points are inside the domain. The first
and the last points, in the azimuthal direction, are superposed in order to facilitate
the boundary condition implementation which is given by
\begin{eqnarray}
	(\rho)_{i,j,K_{MAX}} &=& (\rho)_{i,j,1} \, \mbox{,} \nonumber\\
	(u)_{i,j,K_{MAX}} &=& (u)_{i,j,1} \, \mbox{,} \nonumber\\
	(v)_{i,j,K_{MAX}} &=& (v)_{i,j,1} \, \mbox{,} \\
	(w)_{i,j,K_{MAX}} &=& (w)_{i,j,1} \, \mbox{,} \nonumber\\
	(e)_{i,j,K_{MAX}} &=& (e)_{i,j,1} \, \mbox{.} \nonumber
	\label{eq:periodicity}
\end{eqnarray}

  \section{High Performance Computing}

The current section presents an overview of the LES solver and
discusses the high performance computing implementations introduced
into the code. A study on the parallel performance of JAZzY using 
multiple processors is presented and discussed in the end of the section.

\subsection{Mesh Generation}

The LES solver presents a parallel-IO feature in which each MPI partition 
reads its correspondent portion of the mesh. Therefore, a 3-D grid 
generator is developed in order to provide partitioned CGNS mesh files 
to the LES solver. The CGNS standard \cite{Poirier98,Poirier00,legensky02} 
is build on the HDF5 library \cite{folk11,folk99}. This library is a general 
scientific format adaptable to virtually any scientific or engineering 
application. It provides tools to efficiently read and write data structured 
in a binary tree fashion. This data structure can handle many types of 
queries very efficiently \cite{Bentley-1975,Bentley-1979} such as 
time-dependent CFD solution. 

\begin{figure}[htb!]
       \begin{center}
		   \subfigure[2-D partitioning in the axial and 
		   azimuthal direction.]{
           \includegraphics[width=0.6\textwidth]
		   {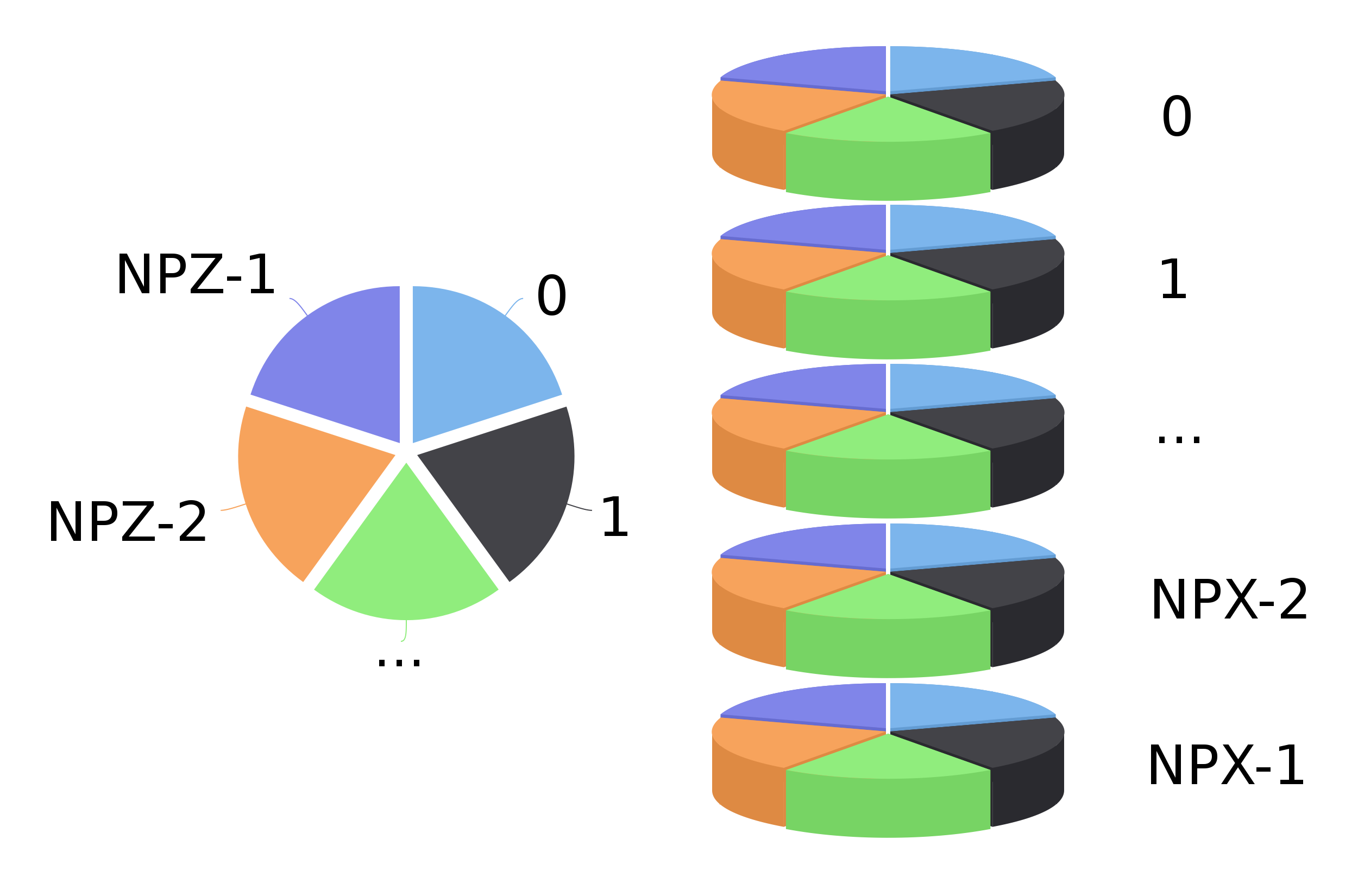} \label{fig:XZ_1}
		   }
		   \subfigure[Mapping of the 2-D partitioning.]{
           \includegraphics[width=0.6\textwidth]
		   {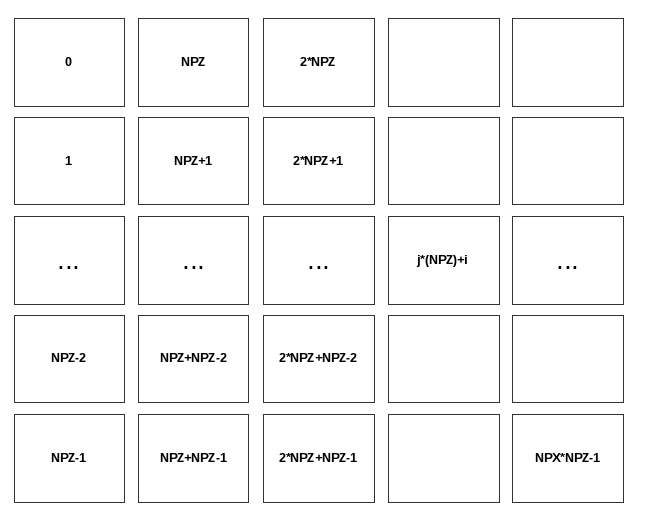} \label{fig:XZ_2}
		   }
		   \caption{2-D partitioning and mapping.}\label{fig:comm_2d}
       \end{center}
\end{figure}

Figure \ref{fig:XZ_1} illustrates the segmentation of the domain into
the axial and azimuthal directions while Fig.\ \ref{fig:XZ_2} presents 
the mapping of the domain. The index of each partition, indicated in Fig.\ 
\ref{fig:XZ_2}, is based on a matrix index system in which the rows 
represent the position in the axial direction and the columns represent the 
position in the azimuthal direction. The partition index starts at zero to 
be consistent with the message passing interface standard. NPX and NPZ denote 
the number of partitions in the axial and azimuthal directions, respectively. 

Table \ref{tab:mesh-gen} presents the algorithm of the mesh generator.
The user can provide geometry and mesh point distribution parameters so 
the grid generator can create a 2-D mesh. A complete 2-D grid, from a 
different mesh generator, can also be provided by the user to the mesh 
generator. In the sequence, the 2-D grid is partitioned in the axial 
direction. After the partitioning in the axial direction, each portion 
of the mesh is extruded in the azimuthal direction respecting the 
positioning of the MPI partitions. Each portion of the mesh is written 
using the CGNS standard. 
\begin{table}[htbp]
\begin{center}
\caption{Mesh generator overview.}
\label{tab:mesh-gen}
\begin{tabular}{|l|lc|}
\hline
 & & \\
1 & BEGIN & \\
2 & Read input data & \\
3 & Read mesh or create 2-D mesh & \\
4 & Perform balanced partitioning in axial direction & \\
5 & Perform balanced partitioning in azimuthal direction & \\
6 & Rotate the mesh partition in the azimuthal direction & \\
7 & Write a CGNS mesh file for each partition & \\
8 & END & \\
\hline
\end{tabular}
\end{center}
\end{table}

\begin{figure}[htb!]
       \begin{center}
           \includegraphics[width=0.5\textwidth]
		   {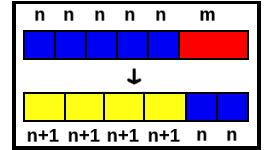}
		   \caption{Balancing procedure performed during the
		   patitioning of the mesh.}\label{fig:balance}
       \end{center}
\end{figure}

The division of the mesh in the axial and azimuthal directions is performed 
towards a well balanced distribution of points. Firstly, the total number of 
grid points in one direction is divided by the number of domains in the same 
direction. The remaining points are spread among the partitions in the case 
which the division is not exact. Figure \ref{fig:balance} illustrates the 
balancing procedure performed in each direction during the partitioning of 
the computational grid.

\subsection{JAZzY Overview}

JAZzY is the LES solver presented in the current work. Table 
\ref{tab:JAZzY} presents a brief overview of JAZzY. In the beginning
of the calculation every MPI partition reads the same ASCII file which 
provides input data such as flow configurations and simulation settings.
In the sequence, each MPI partition reads its correspondent CGNS mesh file. 
The Jacobian and the metric terms are calculated after the I-O procedure. 
Then, each processor sets the initial conditions defined in the input data 
and it performs an asynchronous communication before starting iterations in 
order to solve the compressible LES equations.
\begin{table}[htbp]
\begin{center}
\caption{The JAZzY code overview.}
\label{tab:JAZzY}
\begin{tabular}{|l|lc|}
\hline
 & & \\
1 & Read input data & \\
2 & Read mesh & \\
3 & Calculate Jacobian & \\
4 & Calculate metric terms & \\
5 & Set up initial conditions & \\
6 & Asynchronous communication & \\
7 & WHILE (it. $<$ max nb it.) & \\
  & \; Compute inviscid flux vectors & \\
  & \; Compute artificial dissipation operator & \\
  & \; Calculate inviscid flux contributions to the residue & \\
  & \; Compute viscous flux vectors and SGS viscosity& \\
  & \; Calculate viscous flux contributions to the residue & \\
  & \; Calculate time step & \\
  & \; Perform multi-step explicit time integration & \\
  & \; Update the solution and boundary conditions & \\
8 & END WHILE & \\
9 & Output results & \\
10 & END & \\
\hline
\end{tabular}
\end{center}
\end{table}

The first operation, which is performed in the iteration loop, is the 
computation of the inviscid flux vectors. Then, the artificial 
dissipation operator is calculated. Asynchronous communications are 
performed during this computation. After the data exchange, the inviscid
terms of the LES formulation is calculated using the convective operator 
and the artificial dissipation terms. The viscous terms are calculated 
in sequence and their contributions are added to the right-hand side of 
the LES equations.

The time integration is performed using a five step Runge-Kutta time
integration scheme after the calculation of the numerical fluxes. This 
time marching scheme calculates the inviscid and viscous terms recursively 
through the inner steps. Therefore, multiple communications are performed 
during the time integration. The solution, boundary conditions and fluid 
viscosity are updated after the time marching. Asynchronous communications 
are performed for the periodicity condition. Blocking communications are 
performed in order to calculate properties at the centerline singularity. 
After the updates, neighbor partitions exchange data using non blocking MPI 
routines. The SGS viscosity is calculated in the end of the iteration loop. 
The Vreman and the static Smagorinsky models does not request the use of 
communications. The calculation of the dynamic Smagorinsky SGS viscosity 
is performed using blocking data exchange in order to calculate properties 
on the second level filtering. Finally, when the requested number of 
iterations is achieved, each MPI partition appends the solution to the 
output CGNS file. 

\subsection{Communication}

\begin{figure}[htb!]
       \begin{center}
       {\includegraphics[width=0.8\textwidth]
	   {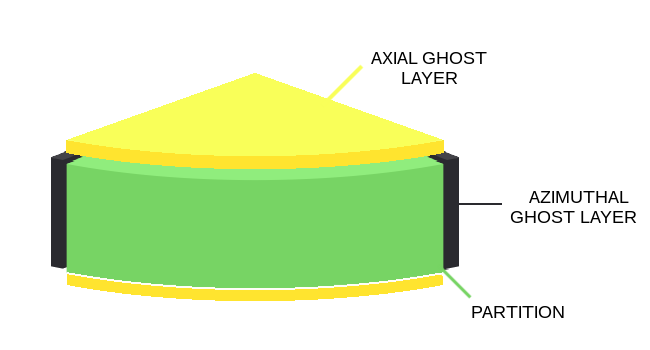}}\\
       \caption{Ghost points creation procedure.}\label{fig:ghost}
       \end{center}
\end{figure}

Numerical data exchange between the partitions is necessary in order to 
perform pa\-ral\-lel computation. 
Ghost points are added to the boundaries of local partition mesh at the 
main flow direction and at the azimuthal direction in order to carry 
information of the neighbor points. The artificial dissipation scheme 
implemented in the code \cite{jameson_mavriplis_86} uses a five points 
stencil which demands information of the two neighbors of a given mesh 
point. Hence, a two layer ghost points is created at the beginning and at 
the end of each partition. Figure \ref{fig:ghost} presents the layer of
ghost points used in the present code. The yellow and black layers 
represent the axial and azimuthal ghost points respectively. The green 
region is the partition mesh.

After the ghost points creation, each processor performs the computation. 
Communication between neighbor partitions are performed in order to allow 
data information pass through the computational domain. Blocking and 
non-blocking communications are used in the present work. In the blocking 
communication approach, the partition which sends the information only 
restarts the computation after the neighbor partition, which is the receiver, 
has finished to read the data. The same does not occur for non-blocking 
communications. The partition which has sent data does not need to wait a 
signal from the receiver. The developer is responsible to assure that data is 
communicated before being accessed through the use of MPI wait functions along 
the code. Most of the data exchange are performed using non-blocking MPI 
communication routines in the current research. Only the centerline boun\-dary 
condition and the communications for the dynamic Smagorinky model 
\cite{Germano91,moin91} are performed using blocking MPI communication 
routines. The first is performed in order to assure reproducibility of the 
solver. The second is performed using blocking communication because it
represents only a small portion of the code. 

\begin{figure}[htb!]
       \begin{center}
       {\includegraphics[width=0.8\textwidth]{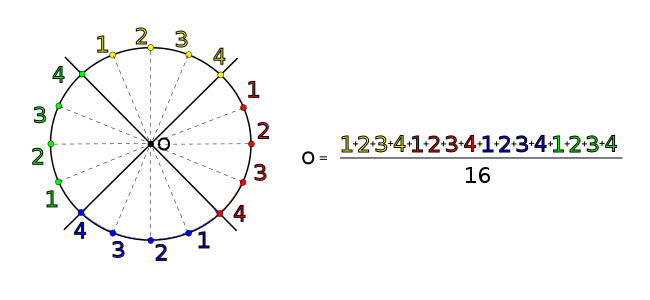}}\\
       \caption{Singularity averaging in parallel.}\label{fig:singularity}
       \end{center}
\end{figure}

The meshes used in the current research have a singularity at the centerline. 
It is necessary to correctly treat this region for the sake of data 
consistency. Therefore, properties are extrapolated to the singularity in 
radial direction and, in the sequence, the master partition collects all 
data from the partitions that share the same singularity point and allocates 
into one single vector. After the allocation, the properties are averaged in a 
sequential fashion and the result is spread to the neighbors in the azimuthal 
direction. Figure \ref{fig:singularity} illustrates the singularity treatment 
for a configuration with 16 points in the azimuthal direction. The yellow, 
red, blue and green colors represent the four partitions in the azimuthal 
direction. Such procedure does not use collective communications in order to 
preserve the commutative property during the averaging. This blocking 
communication is very important in order to achieve the binary reproducibility 
of the computational tool \cite{balaji2013}. The use of such communication is 
motivated by the work of Arteaga {\it et.\ al} \cite{arteaga14}.

\begin{figure}[ht]
       \begin{center}
		   \subfigure[Forward communication between partitions.]{
		   \includegraphics[width=0.475\textwidth]{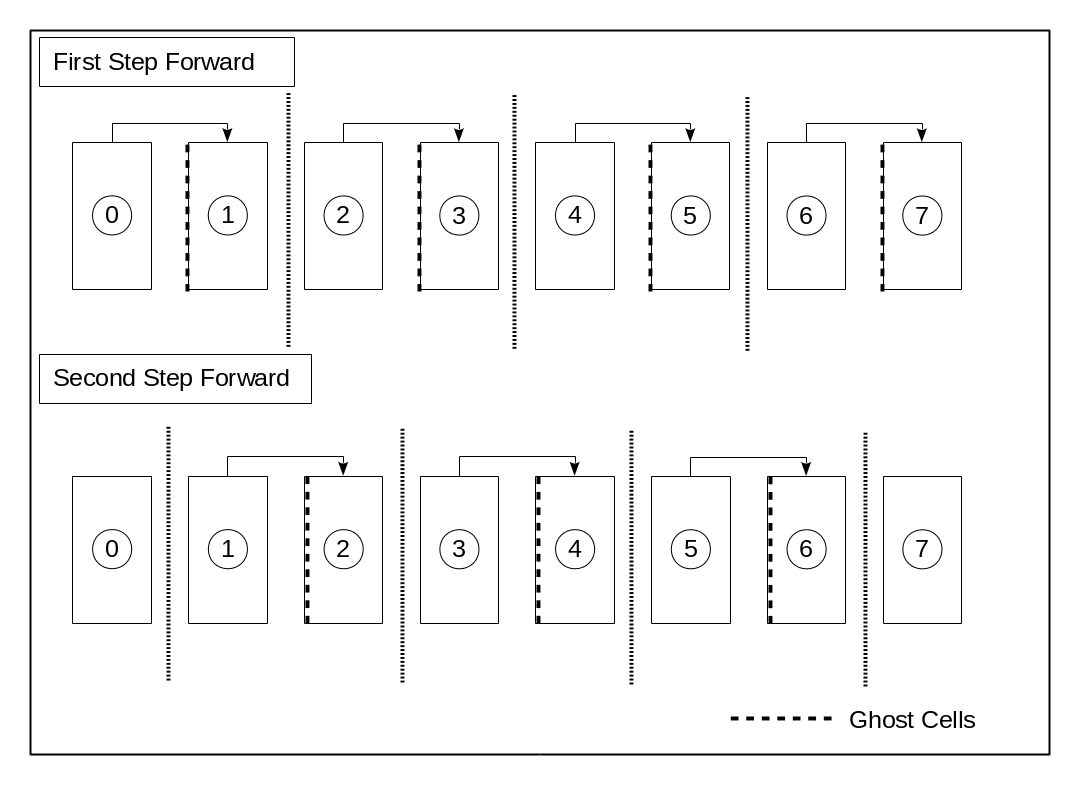}
       	   \label{fig:forward}
		   }
		   \subfigure[Backward Communication between partitions.]{
		   \includegraphics[width=0.475\textwidth]{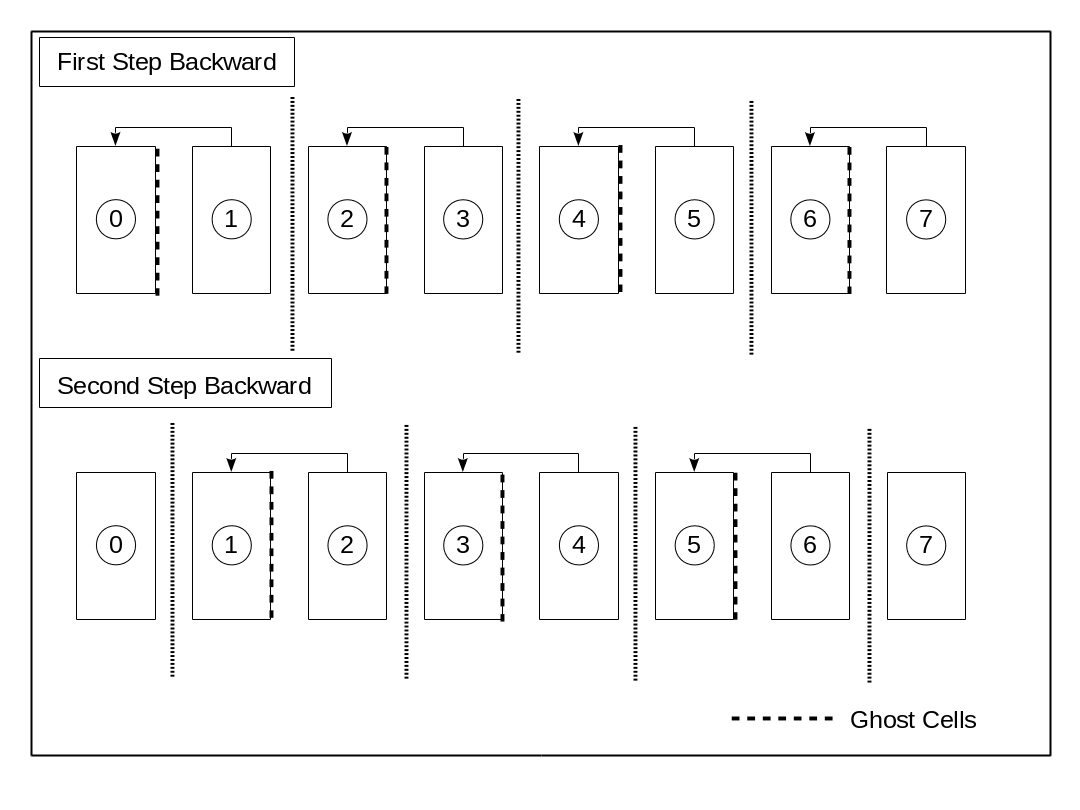}
		   \label{fig:backward}
		   }
		   \caption{Forward and Backward communication schemes used in 
		   order to exchange information between neighbor 
		   partitions}\label{fig:comm-hpc}
       \end{center}
\end{figure}

Non-blocking communication is not available for the communication performed 
by the dynamic Smagorinsky model. Only blocking communication are implemented
because it represents only a small portion of the code. This data exchange 
is firstly performed in the azimuthal direction and in the sequence in the 
axial direction. The azimuthal communication is performed in four blocking 
steps as Figs.\ \ref{fig:forward} and \ref{fig:backward} demonstrate. Initially, 
the communication is performed in the forward direction. Even partitions send 
information of their two last local layers to the ghost points at the left 
of odd partitions. If the last partition is even, it does not share 
information in this step. In the sequence, odd partitions send information 
of their two last local layers to the ghost points at the left of even 
partitions. If the last partition is odd, it does not share information in 
this step. The third and the fourth steps are backward communications. First, 
odd partitions send data of their two first local layers to the ghost points 
at the right of even partitions. Finally, all even partitions, but the first, 
send data of their two first local layers to the ghost points at the right of 
odd partitions. Communication in the axial direction are performed using the 
same approach.

\subsection{Computational Resources}

The current work is included into a national project know as CEPID-CeMEAI
\cite{cepid}. This project provides access to a SGI cluster. The machine 
has 104 computational nodes and each one has two deca-core 2.8 GHz {\it 
Intel Xeon\textsuperscript{\textregistered}} \, E52680v2 processors and 
128 Gb DDR3 1866MHz random access memory. The entire cluster has 2080 
computational cores available for the project members. The storage can be 
performed using the network file system (NFS) or the {\it 
Lustre\textsuperscript{\textregistered}} file system \cite{lustre}. Both 
storage system have 175 Tb available for the users. The network communication 
is performed using Infiniband and Gigabit Ethernet. The Operational system is
the Red Hat Enterprise Linux \cite{redhat} and the job scheduler is the Altair
PBS Pro \cite{pbs}.

The Intel Composer XE Update 2 compiler, version 15.0.2.164 is used in the 
present work. The code is compiled using optimization flags in order to achieve
the best computational performance. The best compilation flags were tested 
in the present work. The flags which provided the best results are:
\begin{itemize}
	\item O3: enables aggressive optimization such as global code scheduling,
		software pipelining, predication and speculation, prefetching, scalar 
		replacement and loop transformations;
	\item xHost: tells the compiler to generate instructions for the highest 
instruction set available on the compilation host processor;
    \item ipo: automatic, multi-step process that allows the compiler to 
		analyze the code and determine where you can benefit from specific 
		optimizations;
	\item no-prec-div: enables optimizations that give slightly less precise 
		results than full division; 
	\item assume buffered\_io : tells the compiler to accumulate records in a 
		buffer;
	\item override-limits: deals with very large, complex functions and loops.
\end{itemize}


\subsection{Computational Performance}

Parallel computation can largely decrease the time of CFD simulations. 
However, the parallel upgrade of a serial solver must be carefully 
performed. The communication between partitions is not free and can 
affect the computational performance in parallel. The partitioning of 
the computational domain increases the number of communication between 
processors. When the number of points of a partitions becomes smaller 
simulation spend less time computing and more time performing 
communications. Consequently, the parallel performance of the solver 
is deteriorated.

The speedup is one of the most common figures for the performance 
evaluation of parallel algorithms and architectures \cite{Ertel94} and 
it is used in the present work in order to measure the computational 
performance of the parallel solver and compare with the ideal case. 
Different approaches are used by the scientific community in order to 
calculate the speedup \cite{gustafson88, xian10}. In the present work
the speedup, ${Sp}_{N}$, is given by 
\begin{equation}
	{Sp}_{N} = \frac{T_{s}}{T_{N}} \, \mbox{.}
	\label{eq:speedup}
\end{equation}
in which $T_{N}$ and $T_{s}$ stand for the time spent to perform one 
thousand iterations using $N$ processors and one single processor, 
respectively. The efficiency as function of the number of processors, 
$\eta_{N}$, is written considering Amdahl's law \cite{amdahl67} as
\begin{equation}
	\eta_{N} = \frac{{Sp}_{N}}{N} \, \mbox{.}
	\label{eq:eff}
\end{equation}

The performance of the solver is evaluated for a large-eddy simulation
of a isothermic perfectly expanded turbulent jet flow without any SGS 
model. The impact of SGS models on the parallel performance of the code 
is evaluated in the work of Junqueira-Junior \cite{Junior16}. Different 
parameters such as mesh size, partitioning configurations and number 
of processors are used to study the parallel behavior of the code in 
the current article. 

Table \ref{tab:mesh} presents the different grid configurations used 
in the current work. There are nine meshes whose total number of points 
doubles every time. The first column presents the name of the mesh. The 
second, third and fourth columns present the number of points in the 
axial, radial and azimuthal directions, respectively. The last column 
indicates the total number of points of the mesh. The grid point distribution 
in the azimuthal direction is fixed and it is equals to 361. The smallest 
grid, named as Mesh A, has approximately 5.9 million points while the biggest 
grid presents approximately 1.0 billion points. 
\begin{table}[htbp]
\begin{center}
\caption{Configuration of computational meshes used in the
	current study.}
\label{tab:mesh}
\begin{tabular}{|c|c|c|c|c|}
\hline
Mesh & No.\ Pt.\ Axial Direct.\ & No.\ Pt.\ Radial Direct.\ & 
No.\ Pt.\ Azimut.\ Direct.\ & No.\ Pt. \\
\hline \hline
	A & 128 & 128 & 361 & 5.9M \\
\hline
    B & 256 & 128 & 361 & 11.8M \\
\hline
	C & 256 & 256 & 361 & 23.7M \\
\hline
	D & 512 & 256 & 361 & 47.3M \\
\hline
	E & 512 & 512 & 361 & 94.6M \\ 
\hline
	F & 1024 & 512 & 361 & 189.3M \\
\hline
	G & 1024 & 1024 & 361 & 378.5M \\
\hline
	H & 2048 & 1024 & 361 & 757.1M \\
\hline
	I & 1700 & 1700 & 361 & 1.0B \\
\hline
\end{tabular}
\end{center}
\end{table}

The strong scalability test is used in the present paper in order to 
evaluate the parallel performance of the code. This test is a measure
of the evolution of speedup and efficiency of a given problem, with a 
fixed size, as the number of processors increases. Simulations are 
performed using up to 400 processors in the present paper. 
Different partitioning configurations are used in order to measure its 
effects on the parallel computational efficiency. Table \ref{tab:part} 
presents the number of partitions in the azimuthal direction for each 
number of processors used to study the scalability of the solver. The 
first column indicates the computational resource while the second 
columns indicates the number of zones in the azimuthal direction used
to evaluate the effects of the partitioning on the computation.
\begin{table}[htbp]
\begin{center}
\caption{Number of partitions in the azimuthal direction
	for a given number of processors.}
\label{tab:part}
\begin{tabular}{|c|c|}
\hline
	No.\ Proc.\ & No.\ of Part.\ in the Azimuthal Dir.\  \\
\hline \hline
	1 & 1 \\
\hline
	2 & \begin{tabular}{c|c}
		1 & 2
	    \end{tabular}\\
\hline
	5 &  \begin{tabular}{c|c}
		1 & 5
	    \end{tabular}\\
\hline
	10 & \begin{tabular}{c|c|c|c}
		1 & 2 & 5 & 10
	    \end{tabular}\\
\hline
	20 & \begin{tabular}{c|c|c|c|c|c}
		1 & 2 & 4 & 5 & 10 & 20
	     \end{tabular}\\
\hline
	40 & \begin{tabular}{c|c|c|c|c|c|c|c}
		1 & 2 & 4 & 5 & 8 & 10 & 20 & 40
         \end{tabular}\\
\hline
	80 & \begin{tabular}{c|c|c|c|c|c|c}
		2 & 4 & 5 & 8 & 10 & 20 & 40
         \end{tabular}\\
\hline
	100 & \begin{tabular}{c|c|c|c|c|c}
		2 & 4 & 5 & 10 & 20 & 25
          \end{tabular}\\
\hline
	200 & \begin{tabular}{c|c|c|c|c|c}
		4 & 8 & 10 & 20 & 25 & 50
          \end{tabular}\\
\hline
	400 & \begin{tabular}{c|c|c|c|c}
		8 & 16 & 20 & 25 & 50
	\end{tabular}\\
\hline
\end{tabular}
\end{center}
\end{table}

Isothermic perfectly expanded jet flow simulations are performed using 
different grid sizes and different partition configurations. The Reynolds 
number of the jet is $1.5744 \times 10^{6}$ for the present simulations and a 
flat-hat profile with Mach number of 1.4 is imposed at the entrance of 
the computational domain. A stagnated flow is used as initial condition 
for the simulations. The time increment is $2.5 \times 10^{-4}$ seconds for the tests 
performed. Numerical results of such configuration using the same code
are presented in the work described in Ref.\ [\citen{Junior15,jr16-aiaa}]. 
In the present article each simulation performs 
1000 iterations or 24 hours of computation. An average of the CPU time 
per iteration through the simulations is measured in order to calculate 
and compare computational cost, speed-up and efficiency of the solver. 

Tables \ref{tab:scala-a} to \ref{tab:scala-i} present, for a given
number of processors, the partitioning configuration which provides
the best averaged CPU time per iteration, its correspondent speedup and
its correspondent computational efficiency for meshes A to I respectively.
One can notice that meshes F,G,H and I are too big and cannot fit into 
one single node of the computational cluster used in the current paper. 
The scalability study started with 40 processors for mesh F, 80 processors 
for mesh G, and 200 processors for mesh H and mesh I. The efficiency of
the solver is considered 100\% at the starting point in order to have a 
reference for the cases in which is not possible to perform a simulation
using one single computational core.
\begin{table}[htbp]
\begin{center}
	\caption{Computational performance of Mesh A.}
\label{tab:scala-a}
\begin{tabular}{|c|c|c|c|c|}
\hline
	No.\ Proc.\ & Av.\ CPU time & Speedup & Efficiency & No.\ Azim.\ Part. \\
\hline \hline
    1 & 2.97E+01 & 1.00E+00 & 1.00E+00 & 1 \\
    2 & 1.14E+01 & 2.61E+00 & 1.31E+00 & 1 \\
    5 & 3.96E+00 & 7.49E+00 & 1.50E+00 & 1 \\
   10 & 2.27E+00 & 1.31E+01 & 1.31E+00 & 2 \\
   20 & 1.57E+00 & 1.89E+01 & 9.43E-01 & 5 \\
   40 & 8.32E-01 & 3.57E+01 & 8.91E-01 & 20\\
   80 & 4.60E-01 & 6.45E+01 & 8.07E-01 & 20\\
  100 & 3.80E-01 & 7.81E+01 & 7.81E-01 & 20\\
  200 & 2.37E-01 & 1.25E+02 & 6.26E-01 & 50\\
  400 & 1.77E-01 & 1.67E+02 & 4.19E-01 & 50\\
\hline
\end{tabular}
\end{center}
\end{table}
\begin{table}[htbp]
\begin{center}
	\caption{Computational performance of Mesh B.}
\label{tab:scala-b}
\begin{tabular}{|c|c|c|c|c|}
\hline
	No.\ Proc.\ & Av.\ CPU time & Speedup & Efficiency & No.\ Azim.\ Part. \\
\hline \hline 
	1 & 5.78E+01 & 1.00E+00 & 1.00E+02 & 1 \\
	2 & 2.60E+01 & 2.23E+00 & 1.11E+02 & 2 \\
	5 & 7.56E+00 & 7.65E+00 & 1.53E+02 & 1 \\
	10 & 4.38E+00 & 1.32E+01 & 1.32E+02 & 2 \\
	20 & 3.09E+00 & 1.87E+01 & 9.36E+01 & 4 \\
	40 & 1.61E+00 & 3.59E+01 & 8.98E+01 & 8 \\
	80 & 8.43E-01 & 6.86E+01 & 8.58E+01 & 10 \\
	100 & 7.25E-01 & 7.98E+01 & 7.98E+01 & 25 \\
	200 & 3.92E-01 & 1.48E+02 & 7.38E+01 & 25 \\
	400 & 2.60E-01 & 2.23E+02 & 5.57E+01 & 25 \\
\hline
\end{tabular}
\end{center}
\end{table}
\begin{table}[htbp]
\begin{center}
	\caption{Computational performance of Mesh C.}
\label{tab:scala-c}
\begin{tabular}{|c|c|c|c|c|}
\hline
	No.\ Proc.\ & Av.\ CPU time & Speedup & Efficiency & No.\ Azim.\ Part. \\
\hline \hline 
1   & 1.18E+02 & 1.00E+00 & 1.00E+02 & 1 \\ 
2   & 5.18E+01 & 2.28E+00 & 1.14E+02 & 1 \\
5   & 1.68E+01 & 7.03E+00 & 1.41E+02 & 1 \\
10  & 9.15E+00 & 1.29E+01 & 1.29E+02 & 2 \\
20  & 6.31E+00 & 1.87E+01 & 9.35E+01 & 4 \\
40  & 3.27E+00 & 3.61E+01 & 9.03E+01 & 8 \\
80  & 1.77E+00 & 6.66E+01 & 8.33E+01 & 10 \\
100 & 1.47E+00 & 8.05E+01 & 8.05E+01 & 20 \\
200 & 7.89E-01 & 1.50E+02 & 7.48E+01 & 25 \\
400 & 5.07E-01 & 2.33E+02 & 5.82E+01 & 50 \\
\hline
\end{tabular}
\end{center}
\end{table}
\begin{table}[htbp]
\begin{center}
	\caption{Computational performance of Mesh D.}
\label{tab:scala-d}
\begin{tabular}{|c|c|c|c|c|}
\hline
	No.\ Proc.\ & Av.\ CPU time & Speedup & Efficiency & No.\ Azim.\ Part. \\
\hline \hline 
1	& 2.67E+02	& 1.00E+00	& 1.00E+02 & 1  \\ 
2	& 1.04E+02	& 2.56E+00	& 1.28E+02 & 1  \\
5	& 3.28E+01	& 8.14E+00	& 1.63E+02 & 1  \\
10	& 1.74E+01	& 1.54E+01	& 1.54E+02 & 2  \\
20	& 1.21E+01	& 2.20E+01	& 1.10E+02 & 4  \\
40	& 6.31E+00	& 4.23E+01	& 1.06E+02 & 8  \\
80	& 3.27E+00	& 8.17E+01	& 1.02E+02 & 8  \\
100	& 2.73E+00	& 9.79E+01	& 9.79E+01 & 20 \\
200	& 1.49E+00	& 1.78E+02	& 8.92E+01 & 20 \\
400	& 8.93E-01	& 2.99E+02	& 7.47E+01 & 25 \\
\hline
\end{tabular}
\end{center}
\end{table}
\begin{table}[htbp]
\begin{center}
	\caption{Computational performance of Mesh E.}
\label{tab:scala-e}
\begin{tabular}{|c|c|c|c|c|}
\hline
	No.\ Proc.\ & Av.\ CPU time & Speedup & Efficiency & No.\ Azim.\ Part. \\
\hline \hline 
1	& 7.02E+02	& 1.00E+00	& 1.00E+02 & 1  \\ 
2	& 2.18E+02	& 3.22E+00	& 1.61E+02 & 1  \\
5	& 7.33E+01	& 9.58E+00	& 1.92E+02 & 1  \\
10	& 3.68E+01	& 1.91E+01	& 1.91E+02 & 2  \\
20	& 2.85E+01	& 2.46E+01	& 1.23E+02 & 2  \\
40	& 1.27E+01	& 5.53E+01	& 1.38E+02 & 8  \\
80	& 6.67E+00	& 1.05E+02	& 1.32E+02 & 2  \\
100	& 5.54E+00	& 1.27E+02	& 1.27E+02 & 20 \\
200	& 3.02E+00	& 2.33E+02	& 1.16E+02 & 20 \\
400	& 1.70E+00	& 4.14E+02	& 1.03E+02 & 16 \\
\hline
\end{tabular}
\end{center}
\end{table}
\begin{table}[htbp]
\begin{center}
	\caption{Computational performance of Mesh F.}
\label{tab:scala-f}
\begin{tabular}{|c|c|c|c|c|}
\hline
	No.\ Proc.\ & Av.\ CPU time & Speedup & Efficiency & No.\ Azim.\ Part. \\
\hline \hline 
40	 & 2.47E+01	& 4.00E+01	& 1.00E+02 & 4  \\
80	 & 1.44E+01	& 6.87E+01	& 8.58E+01 & 8  \\
100	 & 1.04E+01	& 9.54E+01	& 9.54E+01 & 10 \\
200	 & 5.47E+00	& 1.81E+02	& 9.04E+01 & 8  \\
400	 & 3.12E+00	& 3.17E+02	& 7.92E+01 & 8  \\
\hline
\end{tabular}
\end{center}
\end{table}
\begin{table}[htbp]
\begin{center}
	\caption{Computational performance of Mesh G.}
\label{tab:scala-g}
\begin{tabular}{|c|c|c|c|c|}
\hline
	No.\ Proc.\ & Av.\ CPU time & Speedup & Efficiency & No.\ Azim.\ Part. \\
\hline \hline 
80	& 2.61E+01	& 8.00E+01	& 1.00E+02 & 8  \\ 
100	& 2.17E+01	& 9.63E+01	& 9.63E+01 & 10 \\
200	& 1.17E+01	& 1.78E+02	& 8.92E+01 & 8  \\
400	& 6.76E+00	& 3.09E+02	& 7.72E+01 & 20 \\
\hline
\end{tabular}
\end{center}
\end{table}
\begin{table}[htbp]
\begin{center}
	\caption{Computational performance of Mesh H.}
\label{tab:scala-h}
\begin{tabular}{|c|c|c|c|c|}
\hline
	No.\ Proc.\ & Av.\ CPU time & Speedup & Efficiency & No.\ Azim.\ Part. \\
\hline \hline 
200	& 2.13E+01	& 2.00E+02	& 1.00E+02 & 8  \\ 
400	& 1.13E+01	& 3.75E+02	& 9.37E+01 & 16 \\
\hline
\end{tabular}
\end{center}
\end{table}
\begin{table}[htbp]
\begin{center}
	\caption{Computational performance of Mesh I.}
\label{tab:scala-i}
\begin{tabular}{|c|c|c|c|c|}
\hline
	No.\ Proc.\ & Av.\ CPU time & Speedup & Efficiency & No.\ Azim.\ Part. \\
\hline \hline 
200	& 4.09E+01	& 2.00E+02	& 1.00E+02 & 4  \\ 
400	& 1.52E+01	& 5.39E+02	& 1.35E+02 & 16 \\
\hline
\end{tabular}
\end{center}
\end{table}

\begin{figure}[htb!]
       \begin{center}
           \includegraphics[width=0.7\textwidth]
		   {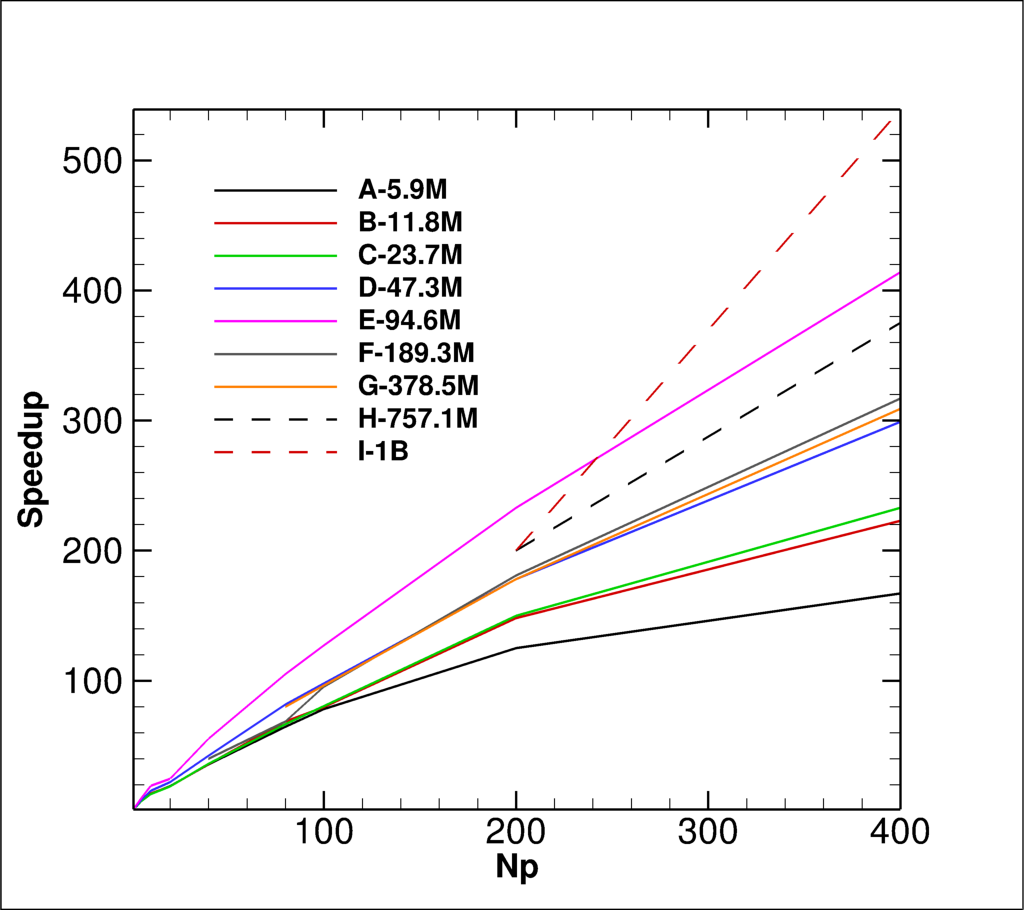}
		   \caption{Speedup curve of the LES solver for 
		   different mesh sizes.}\label{fig:speedup}
       \end{center}
\end{figure}

\begin{figure}[htb!]
       \begin{center}
           \includegraphics[width=0.7\textwidth]
		   {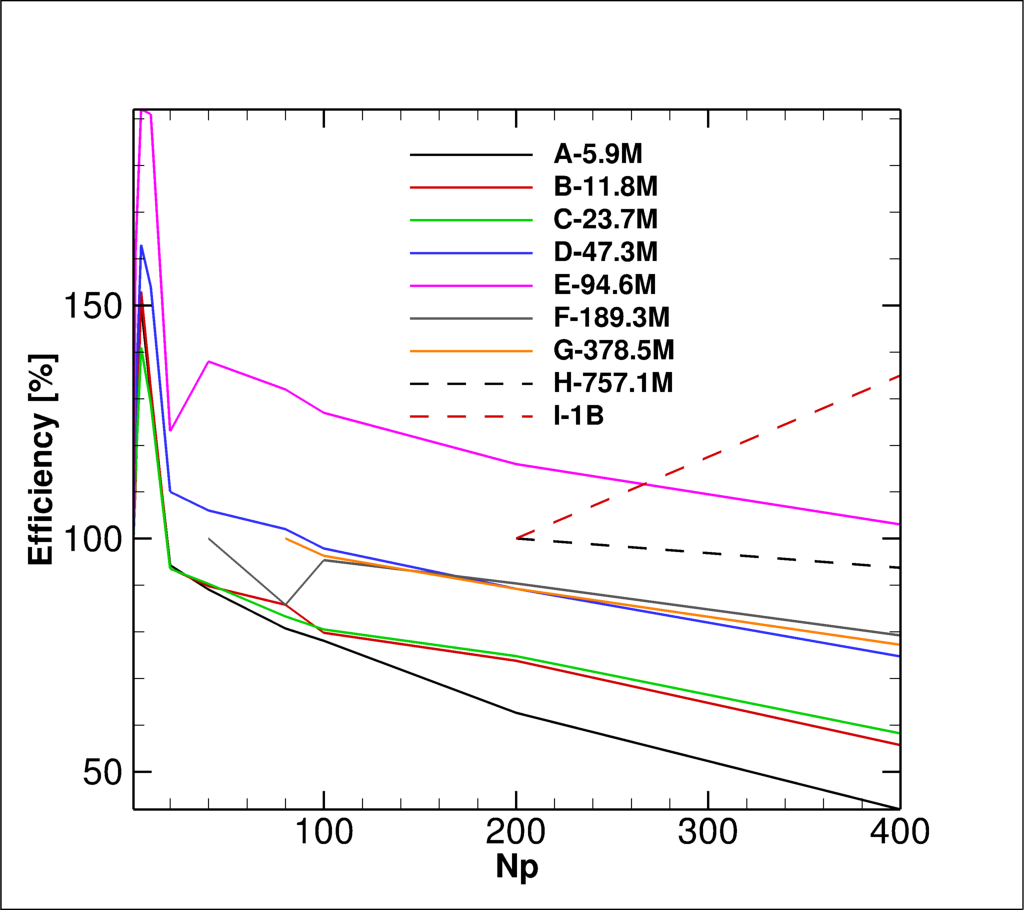}
		   \caption{Parallel efficiency of the LES solver for 
		   different mesh sizes.}\label{fig:efficiency}
       \end{center}
\end{figure}

The evolution of speedup and efficiency as function of the number of
processors for all computational grids used in the current work are
presented in Figs.\ \ref{fig:speedup} and \ref{fig:efficiency}. The
code presents a good scalability, with an efficiency bigger than 75\%, 
for grids which have more than 50 million points. Mesh E presented 
an efficiency of $\approx$ 100\% and speedup of 400 when running on 400 
processors. Such performance is equivalent to the theoretical 
speedup. One can notice a super linear scalability for the cases which 
the speedup reference is the sequential computation and also for mesh 
I. The first can be explained by the fact that there is not a serial 
version of the solver. There is only a parallel version which can run 
simulations using a single computational core. Moreover, cache memory 
can be the bottleneck of a simulation using a given mesh and a given 
number of processors \cite{benzi09}. Such limitation can explain the 
super linear speedup of mesh I. The bottleneck can deteriorate the 
performance of the solver. When the number of processors is increased 
and mesh size conserved, the cache memory can become no longer a 
limitation. This effect can generate super-scalability which can be 
interpreted as computational efficiency greater than 100\%.

\FloatBarrier

Increasing the size of a computational problem can generate a better 
scalability study. The time spent with computation becomes more 
significant when compared to the time spent with communication with the 
growth of a problem. One can notice such effect for meshes A, B, C, D 
and E. The speedup and the efficiency increase with with the growth of 
the mesh size. However, such scalability improvement does not happen
from mesh E to meshes F, G, H and I. This behavior is originated because 
the reference used to calculate speedup and efficiency is not the same 
for all grid configurations. The studies performed using meshes F, G, H and 
I does not use the serial computation as a reference, which is not the 
case for the scalability studies performed in the current paper using 
mesh A, B, C, D and E.

\section{Concluding Remarks}

The current work is a computational performance study of a large eddy 
simulation solver for supersonic jet flow configurations. Nine strong
scalability studies are performed using meshes whose size grows from 
approximately 5.9 million points to approximately 1.0 billion points. 
Different partitioning configurations are used to evaluate its effects
on the computational performance of the solver. Simulations are run 
on one processor up to 400 computational cores. The speedup and the 
computational efficiency are calculated for every study performed in 
the present article.

The filtered compressible large eddy simulation formulation is written
using a finite-difference centered second-order spatial discretization 
with the explicit addition of artificial dissipation. The time 
integration is performed using a five-steps second order Runge-Kutta 
scheme. Three subgrid scale models are implemented into the solver. 
Message passing interface protocols are used in order to perform the 
computation in parallel. The code presents parallel-IO features. Each 
MPI partition reads its portion of the mesh. A mesh generator is 
created in order to provide balanced CGNS mesh partitions. The solver 
creates two layers of ghost points in the axial and in azimuthal 
direction for each partition in order to exchange data with neighbor 
zones. Communication between partitions are performed using non blocking 
data exchange towards the best computational performance.

The code presented a good scalability for the calculations run in the 
current paper. The averaged CPU time per iteration decays with the 
increase of number of processors in parallel for all computation 
performed by the large eddy simulation solver evaluated in the present
work. Meshes with more than 50 million points indicated an efficiency 
greater than 75\%. The problem with approximately 100 million points 
presented speedup of 400 and efficiency of 100\% when running on 400 
computational cores in parallel. Such performance is equivalent to 
theoretical behavior in parallel. 
It is important to remark the ability of the parallel
solver to treat very dense meshes as the one tested in the present 
paper with approximately 1.0 billion points. Large eddy simulation 
demand very refined grids in order to have a well representation of 
the physical problem of interest. Therefore, it is important to perform 
simulations of such configuration with a good computation efficiency.
One can notice the presence of super-linear speedup in the current study. 
Such behavior can be explained by cache limitations when running simulations 
with low amount of computational resources. Moreover, there is no serial 
version of the code. The sequential study is a parallel version running on 
one single processor.




\FloatBarrier




\section*{Acknowledgments}

The authors gratefully acknowledge the partial support for this research 
provided by Conselho Nacional de Desenvolvimento Cient\'{\i}fico e 
Tecnol\'{o}gico, CNPq, under the Research Grants No.\ 309985/2013-7, No.\ 
400844/2014-1, No.\ 443839/2014-0 and No.\ 150551/2017-1\@. The authors are 
also indebted to the partial financial support received from Funda\c{c}\~{a}o 
de Amparo \`{a} Pesquisa do Estado de S\~{a}o Paulo, FAPESP, under the Research 
Grants No.\ 2013/07375-0 and No.\ 2013/21535-0. 

\newpage



\bibliography{sources/references}

\begin{thebibliography}{10}
\newcommand{\enquote}[1]{``#1''}

\bibitem{Junior16}
Junqueira-Junior, C.~A., {\em Development of a Parallel Solver for Large Eddy
  Simulation of Supersonic Jet Flow\/}, Ph.D. thesis, Instituto Tecno\'{o}gico
  de Aeron\'{a}utica, S\~{a}o Jos\'{e} dos Campos, SP, Brazil, 2016.

\bibitem{jr16-aiaa}
Junqueira-Junior, C., Yamouni, S., {a}o Luiz F.~Azevedo, J., and Wolf, W.~R.,
  \enquote{{Influence of Different Subgrid Scale Models in LES of Supersonic
  Jet Flows},} {\em {\em AIAA Paper No.\ 2016-4093}, 46th AIAA Fuid Dynamics
  Conference, AIAA Aviation Forum\/}, Washington, D.C., Jun. 2016.

\bibitem{Wolf2012}
Wolf, W.~R., Azevedo, J. L.~F., and Lele, S.~K., \enquote{Convective Effects
  and the Role of Quadrupole Sources for Aerofoil Aeroacoustics,} {\em Journal
  of Fluid Mechanics\/}, Vol.~708, 2012, pp.~502--538.

\bibitem{Vreman1995}
Vreman, A.~W., {\em Direct and Large-Eddy Simulation of the Comperssible
  Turbulent Mixing Layer\/}, Ph.D. thesis, Universiteit Twente, 1995.

\bibitem{Mendez10}
Mendez, S., Shoeybi, M., Sharma, A., Ham, F.~E., Lele, S.~K., and Moin, P.,
  \enquote{Large-Eddy Simulations of Perfectly-Expanded Supersonic Jets:
  Quality Assessment and Validation,} {\em {\em AIAA Paper No.\ 2010--0271}\/},
  January 2010.

\bibitem{bridges2008turbulence}
Bridges, J. and Wernet, M.~P., \enquote{Turbulence Associated with Broadband
  Shock Noise in Hot Jets,} {\em AIAA paper\/}, Vol.~2834, 2008, pp.~2008.

\bibitem{folk99}
Folk, M., Cheng, A., and Yates, K., \enquote{HDF5: A File Format and I/O
  Library for High Performance Computing Applications,} {\em Proceedings of
  Supercomputing\/}, Vol.~99, 1999, pp. 5--33.

\bibitem{folk11}
Folk, M., Heber, G., Koziol, Q., Pourmal, E., and Robinson, D., \enquote{An
  Overview of the HDF5 Technology Suite and its Applications,} {\em Proceedings
  of the EDBT/ICDT 2011 Workshop on Array Databases\/}, ACM, 2011, pp. 36--47.

\bibitem{Poirier98}
Poirier, D. and Enomoto, F.~Y., \enquote{The CGNS System,} {\em {\em AIAA Paper
  No.\ 98-3007}, Proceedings of 29th AIAA Fluid Dynamics Conference\/},
  Albuquerque, NM, June 1998.

\bibitem{Poirier00}
Poirier, D. M.~A., Bush, R.~H., Cosner, R.~R., Rumsey, C.~L., and McCarthy,
  D.~R., \enquote{Advances in the CGNS Database Standard for Aerodynamics and
  CFD,} {\em {\em AIAA Paper No.\ 2000-0681}, 38th AIAA Aerospace Sciences
  Meeting \& Exhibit\/}, Reno, NV, Jan. 2000.

\bibitem{legensky02}
Legensky, S.~M., Edwards, D.~E., Bush, R.~H., and Poirier, D., \enquote{CFD
  General Notation System (CGNS) - Status and Future Directions,} {\em {\em
  AIAA Paper No.\ 2002-0752}, Proceedings of 40th AIAA Aerospace Sciences
  Meeting \& Exhibit\/}, Reno, NV, Jan. 2002.

\bibitem{Dongarra95}
Dongarra, J.~J., Otto, S.~W., Snir, M., and Walker, D., \enquote{{An
  Introduction to the MPI Standard},} Tech. rep., Knoxville, TN, USA, 1995.

\bibitem{Sagaut05}
Sagaut, P., {\em Large Eddy Simulation for Incompressible Flows\/}, Springer,
  2002.

\bibitem{Smagorinsky63}
Smagorinsky, J., \enquote{General Circulation Experiments with the Primitive
  Equations: I. The Basic Experiment,} {\em Monthly Weather Review\/}, Vol.~91,
  No.~3, March 1963, pp.~99--164.

\bibitem{Lilly67}
Lilly, D.~K., \enquote{The Representation of Small-Scale Turbulence in
  Numerical Simulation Experiments,} {\em {\em IBM Form No. 320-1951},
  Proceedings of the IBM Scientific Computing Symposium on Environmental
  Sciences\/}, Yorktown Heights, N.Y., 1967, pp. 195--210.

\bibitem{Garnier09}
Garnier, E., Adams, N., and Sagaut, P., {\em Large Eddy Simulation for
  Compressible Flows\/}, Springer, 2009.

\bibitem{vreman2004}
Vreman, A.~W., \enquote{An Eddy-Viscosity Subgrid-Scale Model for Turbulent
  Shear Flow: Algebraic Theory and Applications,} {\em Physics of Fluids\/},
  Vol.~16, No.~10, October 2004.

\bibitem{Lilly65}
Lilly, D.~K., \enquote{On the Computational Stability of Numerical Solutions of
  Time- Dependent Non-Linear Geophysical Fluid Dynamics Problems,} {\em Monthly
  Weather Review\/}, Vol.~93, No.~1, January 1965, pp.~11--25.

\bibitem{Deardorff70}
Deardorff, J.~W., \enquote{A Numerical Study of Three-Dimensional Turbulent
  Channel Flow at Large Reynolds Numbers,} {\em Journal of Fluid Mechanics\/},
  Vol.~41, part 2, 1970, pp.~453--480.

\bibitem{Leonard74}
Leonard, A., \enquote{{Energy Cascade in Large Eddy Simulations of Turbulent
  Fluid Flows},} {\em Adv. Geophys.\/}, Vol.~A18, 1974, pp.~237--48.

\bibitem{Clark79}
Clark, R.~A., Ferziger, J.~Z., and Reynolds, W.~C., \enquote{Evaluation of
  Subgrid-Scale Models Using an Accurately Simulated Turbulent Flow,} {\em
  Journal of Fluid Mechanics\/}, Vol.~91, 1979, pp.~1--16.

\bibitem{vreman1996}
Vreman, B., Geurts, B., and Kuerten, H., \enquote{Large-Eddy Simulation of the
  Turbulent Mixing Layer Using the {C}larck Model,} {\em Theoretical
  Computational Fluid Dynamics\/}, Vol.~8, No.~4, 1996, pp.~309--324.

\bibitem{germano90}
Germano, M., \enquote{Averaging Invariance of the Turbulent Equations and
  Similar Subgrid Scale Modeling,} {\em Center for Turbulence Research
  Manuscript 116\/}, Stanford University and NASA - Ames Research Center, 1990.

\bibitem{moin91}
Moin, P., Squires, K., Cabot, W., and Lee, S., \enquote{A Dynamic Subgrid-Scale
  Model for Compressible Turbulence and Scalar Transport,} {\em Physics of
  Fluids A: Fluid Dynamics (1989-1993)\/}, Vol.~3, No.~11, 1991,
  pp.~2746--2757.

\bibitem{Yoshizawa86}
Yoshizawa, A., \enquote{Statistical Theory for Compressible Turbulent Shear
  Flows, with the Application to Subgrid Modeling,} {\em Physics of Fluids\/},
  Vol.~29, No.~7, July 1986.

\bibitem{BIGA02}
Bigarella, E. D.~V., {\em Three-Dimensional Turbulent Flow Over Aerospace
  Configurations\/}, {M.Sc.} {T}hesis, Instituto Tecnol\'{o}gico de
  Aeron\'{a}utica, S\~ao Jos\'e dos Campos, SP, Brasil, 2002.

\bibitem{Turkel_Vatsa_1994}
Turkel, E. and Vatsa, V.~N., \enquote{{Effect of Artificial Viscosity on
  Three-Dimensional Flow Solutions},} {\em AIAA Journal\/}, Vol.~32, No.~1,
  1994, pp.~39--45.

\bibitem{jameson_mavriplis_86}
Jameson, A. and Mavriplis, D., \enquote{Finite Volume Solution of the
  Two-Dimensional Euler Equations on a Regular Triangular Mesh,} {\em AIAA
  Journal\/}, Vol.~24, No.~4, Apr. 1986, pp.~611--618.

\bibitem{Jameson81}
Jameson, A., Schmidt, W., and Turkel, E., \enquote{Numerical Solutions of the
  Euler Equations by Finite Volume Methods Using Runge-Kutta Time-Stepping
  Schemes,} {\em {\em AIAA Paper 81--1259}, Proceedings of the {AIAA} 14th
  Fluid and Plasma Dynamic Conference\/}, Palo Alto, Californa, USA, June 1981.

\bibitem{Long91}
Long, L.~N., Khan, M., and Sharp, H.~T., \enquote{{A Massively Parallel
  Three-Dimensional Euler/Navier-Stokes Method},} {\em AIAA Journal\/},
  Vol.~29, No.~5, 1991, pp.~657--666.

\bibitem{Bentley-1975}
Bentley, J.~L., \enquote{Multidimensional Binary Search Trees Used for
  Associative Searching,} {\em Communications of the ACM\/}, Vol.~18, No.~9, 9
  1975, pp.~509--517.

\bibitem{Bentley-1979}
Bentley, J., \enquote{Multidimensional Binary Search Trees in Database
  Applications,} {\em IEEE Transactions on Software Engineering\/}, Vol.~SE-5,
  No.~4, 1979, pp.~0--340.

\bibitem{Germano91}
Germano, M., Piomelli, U., Moin, P., and Cabot, W.~H., \enquote{A Dynamic
  Subgridscale Eddy Viscosity Model,} {\em Physics of Fluids A: Fluid
  Dynamics\/}, Vol.~3, No.~7, July 1991.

\bibitem{balaji2013}
Balaji, P. and Kimpe, D., \enquote{On the Reproducibility of MPI Reduction
  Operations,} {\em 2013 IEEE 10th International Conference on High Performance
  Computing and Communications \& IEEE International Conference on Embedded and
  Ubiquitous Computing (HPCC\_EUC)\/}, IEEE, 2013, pp. 407--414.

\bibitem{arteaga14}
Arteaga, A., Fuhrer, O., and Hoefler, T., \enquote{Desingning a
  Bit-Reproducible Portable High-Performance Applications,} {\em Parallel and
  Distributed Processing Symposium, 2014 {IEEE} 28th International\/}, Phoenix,
  AZ, USA, May 2014, pp. 1235--1244.

\bibitem{cepid}
{CEPID-CeMEAI}, \enquote{{C}entro de {C}i\^{e}ncias {A}plicadas a
  Ind\'{u}stria. http://www.cemeai.icmc.usp.br/,} .

\bibitem{lustre}
{Lustre \textregistered}, \enquote{http://www.lustre.org/,} .

\bibitem{redhat}
{RedHat}, \enquote{http://www.redhat.com/,} .

\bibitem{pbs}
{Altair - PBS Works\textsuperscript{TM}}, \enquote{http://www.pbsworks.com/,} .

\bibitem{Ertel94}
Ertel, W., \enquote{On the Definition of Speedup,} {\em {PARLE'94} Parallel
  Architectures and Languages Europe\/}, Springer, Berlin, 1994, pp. 289--300.

\bibitem{gustafson88}
Gustafson, J.~L., \enquote{Reevaluating Amdahl's Law,} {\em Communications of
  the ACM\/}, Vol.~31, No.~5, 1988, pp.~532--533.

\bibitem{xian10}
Sun, X.-H. and Chen, Y., \enquote{Reevaluating Amdahl's Law in Multicore Era,}
  {\em J. Parallel Distrib. Comput.\/}, Vol.~70, No.~2, Feb. 2010,
  pp.~183--188.

\bibitem{amdahl67}
Amdahl, G.~M., \enquote{Validity of the Single Processor Approach to Achieving
  Large Scale Computing Capabilities,} {\em {AFIPS} Conference Proceedings\/},
  Vol.~30, ACM, Atlantic City, N.J., USA, Apr. 1967, pp. 483--485.

\bibitem{Junior15}
Junqueira-Junior, C., Yamouni, S., Azevedo, J. L.~F., and Wolf, W.~R.,
  \enquote{Large Eddy Simulations of Supersonic Jet Flows for Aeroacoustic
  Applications,} {\em {\em AIAA Paper No.\ 2015-3306,} Proceedings of the 33rd
  AIAA Applied Aerodynamics Conference\/}, Dallas, TX, June 2015.

\bibitem{benzi09}
Benzi, J. and Damodaran, M., \enquote{Parallel Three Dimensional Direct
  Simulation Monte Carlo for Simulating Micro Flows,} {\em Parallel
  Computational Fluid Dynamics 2007\/}, Springer, 2009, pp. 91--98.

\end{thebibliography}
\bibliographystyle{aiaa}

\end{document}